\documentclass[12pt,preprint]{aastex}
 \usepackage{epstopdf}
 \usepackage{ifpdf}
 \usepackage{epsfig}
  \usepackage[latin1]{inputenc}
 \usepackage{amsfonts}
 \usepackage{amsthm}
 \usepackage{amssymb, latexsym, amsmath}
 \usepackage{amsbsy}
\usepackage{amstext}
\usepackage{amsopn}
\usepackage{amsgen}
\usepackage[dvips]{color}
 \usepackage{graphicx}
  \usepackage{color}

 \newcommand{\inc}{{\it i}}

 \newcommand{\be}{\begin{equation}}
 \newcommand{\ee}{\end{equation}}
 \newcommand{\ba}{\begin{eqnarray}}
 \newcommand{\ea}{\end{eqnarray}}

 \newcommand{\eb}{\begin{equation}}

\shorttitle{Dynamical evolution of GJ 581d}
\shortauthors{Makarov et al.}
\begin{document}

 \title{
          ${{~~~~~~~~~~~~~~~~~~~~~~}^{{{
         Published~in:~~the~
                 Astrophysical~Journal
        \,,~~Vol.~{\bf{761}}\,,~~article~id.~83~~(December~2012).
                  }}}}$\\
                  ~\\
 Dynamical evolution and spin-orbit resonances of potentially habitable exoplanets. The case of GJ 581d}
 \author{~\\
  {
   \large
  {Valeri V. Makarov}}\\
  {\small{US Naval Observatory, Washington DC 20392}}\\
 {\small{e-mail: ~vvm @ usno.navy.mil~~}},\\
 ~\\
 ~\\
   {
    \large
   {Ciprian Berghea}}\\
  {\small{US Naval Observatory, Washington DC 20392}}\\
 {\small{e-mail: ~ciprian.berghea @ usno.navy.mil~}}~,\\
      ~\\
~\\
   {
   \large
   {Michael Efroimsky}}\\
  {\small{US Naval Observatory, Washington DC 20392}}\\
 {\small{e-mail: ~michael.efroimsky @ usno.navy.mil~}}\\
  }

 \begin{abstract}

\noindent
 GJ 581d is a potentially habitable super-Earth in the multiple system of exoplanets orbiting a nearby M dwarf. We explore this
 planet's long-term dynamics, with an emphasis on its probable final rotation states acquired via tidal interaction with the host.
 The published radial velocities for the star are re-analysed with a benchmark planet detection algorithm, to confirm that there is no
 evidence for the recently proposed two additional planets (f and g). Limiting the scope to the four originally detected planets, we
 assess the dynamical stability of the system and find bounded chaos in the orbital motion. For the planet
 $\,$d$\,$, the characteristic
 Lyapunov time is 38 yr. Long-term numerical integration reveals that the system of four planets is stable, with the eccentricity of
 the planet d changing quasi-periodically in a tight range around 0.27, and with its semimajor axis varying only a little.
 The spin-orbit interaction of GJ 581d with its host star is dominated by the tides exerted by the star on the planet. We model this
 interaction, assuming a terrestrial composition of the mantle. Besides the triaxiality-caused torque and the secular part of the tidal torque, which are
conventionally included into the equation of motion, we also include the tidal torques' oscillating components. It turns out that, dependent on the mantle temperature,
 the planet gets trapped into the 2:1 or an even higher spin-orbit resonance. It is very improbable that the planet could
 have reached the 1:1 resonance. This enhances the possibility of the planet being suitable for sustained life

 \end{abstract}


 \section{Introduction}
 \label{firstpage}
 \label{intro.sec}

 Habitability of an emerging class of super-Earths -- exoplanets with masses greater than that of the Earth but lower than that of Uranus --
 depends on a combination of physical parameters. Among these, crucial is the intensity of irradiation from the host star. A favourable
 rate of irradiation permits water to be available in the liquid form. The irradiation intensity depends on the luminosity of
 the star, the size of the planet's orbit and, to a lesser degree, on the orbital eccentricity.  The chemical composition of the planet's
 atmosphere influences the average temperature and the temperature variations on the surface. For example, estimates show that a certain
 minimal amount of CO$_2$ in the atmosphere of GJ 581d is required to keep water above the freezing point on the surface, while GJ 581c
 is likely to have experienced a runaway greenhouse event \citep{vonP,hu}. Planet GJ 581d, which is the main target of our study, may be
 located on the outer edge of the habitable zone, according to \citet{vonB}.

 We begin our study with addressing the still controversial problem of composition of this planetary system. In Section \ref{orb.sec}, we
 confirm that only the four originally detected planets (b through e) are real and detectable at the 0.99 confidence level by our fully
 automated, fast grid search algorithm. The two additional planets, f and g, proposed by \citet{vog} are not found with any combination
 of the published radial velocity (RV) data.

 The orbit estimation technique employed in our paper is designed, in particular, to produce accurate and robust results for the
 eccentricity of each detected planet -- an important asset for the subsequent dynamical simulations. The dynamical stability and
 the presence of chaos in the orbital motion of the four planets is investigated in Section \ref{cha.sec} by long-term integrations,
 assuming zero inclination and coplanar orbits. The system is found to be long-term stable but strongly chaotic, with the eccentricity
 and semimajor axis of planet d little varying over gigayears. This allows us to investigate, in Section \ref{tid.sec}, the spin-orbit
 dynamics of GJ 581d with its orbital parameters fixed. The planet is assumed to have a terrestrial rheology and a near-zero
 obliquity.

\section{How many planets have been detected in the GJ 581 system?}
\label{orb.sec}

 Although this question has attracted a lot of attention in the literature, some doubts seem to be lingering, and a few papers have been published on subtle features of the planets that may well be nonexistent. The first four planets were discovered with the same instrument, HARPS at La Silla Observatory: GJ 581b \citep{bon}, GJ 581c and d \citep{udr}, and GJ 581e \citep{may}. The combined series of observations with HARPS, published in the latter paper, spanned 1570 days, while a typical single measurement precision was about 1 m s$^{-1}\,$.

 At the first stage of our investigation, we process the HARPS data with a simple benchmark planet detection algorithm. This algorithm is an iterative process of optimisation, based on a simple grid search in the space of free parameters. The grid search is applied only to fitting of the nonlinear parameters, i.e., of the orbital period, eccentricity, and periastron time. The other parameters are determined by a direct Least Squares solution. At the first stage of fitting, the most significant sinusoidal variation in the radial velocity data is determined by a corrected Lomb-Scargle periodogram method \citep{lom,sca}. Our upgrades of the method concern mainly the way the common offset of the RV points is treated. Instead of the traditional subtraction of the mean RV from the data prior to the periodogram computation, we fit the entire set of model functions $[1,\;\cos(2\pi t_j/p_i),\;\sin(2\pi t_j/p_i)]$ for each trial period $p_i$, where $t_j$ are the
 times of observations. This, seemingly trivial modification allows us to update the amplitudes of already detected planets once a new planet is detected, and to keep the original RV measurements unchanged throughout the data reduction cycle.
 The mean RV is inevitably biased when a non-integer number of waves is present in the data. This way, subtraction of the mean RV from the
 data is legitimate only when no periodic signals are present in the data, which defeats the purpose of periodogram analysis.
 It can be proven that the periodogram value in the classical Lomb's algorithm  is equivalent to the absolute change of the $\chi^2$ before
 and after the Least Squares fit of the $\cos$- and $\sin$-terms. However, fitting these terms to the data corrupted by the subtraction of the
 mean RV gives rise to sidelobes and biases in the periodogram. The downside of this modification is that a direct computation of the
 post-fit $\chi^2$ by Scargle's formula is no longer possible. Instead, the complete Least Squares solution of the matrix equation should be
 computed for each trial period. This complication, however, becomes barely noticeable, taken the power of the present-day computers.

 Once the period of the most prominent sinusoidal variation gets determined, a separate grid search is implemented to determine the
 best-fitting eccentricity. This technique is based on the theory of harmonic decomposition of orbital motion \citep[e.g.,][]{bro,kon}.
 The observed radial velocity variation due to the orbital motion of a single planet can be written as
 \eb
 \dot{z}\;=\;-\;C\;\frac{2\pi}{P}\;\sum_{l=1}^\infty \,l\;\alpha_l(e)\;\cos (l\,{\cal{M}}) \;+\;H\;\frac{2\pi}{P}\;\sqrt{1-e^2}\;
 \sum_{l=1}^\infty \,l\;\phi_l(e)\;\sin (l\,{\cal{M}})\;\,,
 \label{AIFI.eq}
 \ee
 where $e$ is the eccentricity, $C$ and $H$ are the Thielle-Innes constants, $P$ is the orbital period determined from the periodogram,
 ${\cal{M}}$ is the mean anomaly, while $\alpha_l(e)$ and $\phi_l(e)$ are coefficients which depend only on $e$ and define the relative
 magnitudes of harmonics. Although the coefficients can be computed via the Bessel functions of the first kind, we find it practical to
 compute them numerically using Kepler's equation at the required resolution in $e$ between 0 and 1, and to store the resulting values of
 the coefficients in a table. The further processing is then reduced to solving the overdetermined equations
 \eb
 \left[\,~1~\quad~\sum_l\,p_l\,\sin (l\,{\cal{M}}_j)~\quad~\sum_l\,q_l\,\cos (l\,{\cal{M}}_j)~\,\right]\;{\bf s}\;=\;\dot z_j~\,,
 \ee
 where $\,p_l\,=\,l\,\alpha_l(e)\,$ and $\,q_l\,=\,l\,\phi_l(e)\,\sqrt{1-e^2}\,$ are tabulated harmonic coefficients, while the integer
 $\,j\,=\,1\,,\,2,\,\ldots\,,\,N\,$ serves to number the available data points. Least-squares solution of the problem is performed for a
 grid of $\,e\,$, and results in selection of the value minimising the post-fit $\,\chi^2\,$. The solution vector $\,\bf s\,$ includes
 three elements: $\,s(1)\,$ is the RV offset, while $\,s(2)\,$, $\,s(3)\,$ are the fitting coefficients from which the Thielle-Innes
 constants $\,C\,$ and $\,H\,$ can be derived. In parallel, a separate grid search is carried out for the phase of orbital motion, i.e.,
 for the mean anomaly at the first observation.

 Finally, we have to estimate the confidence of the orbital solution. We are using an adaptation of the significance F-test, which was
 carefully tested and verified by Monte-Carlo simulations on a large number of real and artificial planetary systems. In this case, the
 null hypothesis for the F-test is that the data set contains a constant offset and a random, uncorrelated noise. Fitting of a harmonic
 $\,[\cos(2\pi t_j/p_i)\,,\;\sin(2\pi t_j/p_i)]\,$ to this set by the least-squares method will generate a $\,\chi^2\,$ change
 corresponding to a random realisation of the F-statistic, $\,F_i\,$, which is distributed as $\,F_{\rm PDF}\,$. The confidence of a
 detection at $\,p_i\,$ is simply the probability of $\,F\le F_i\,$, which is defined by the known cumulative distribution function
 $\,F_{\rm CDF}\,$. It is customary to accept a detection if the confidence is greater than 0.99, i.e., if the observed reduction in
 $\,\chi^2\,$ can happen within the null hypothesis in less than 1 trial out of 100. The confidence of detection is computed as
 \eb
 P_{\rm conf}\;=\;1\;-\;\left[\,1\;-\;F_{\rm CDF}\left(\,\frac{\chi_1^2-\chi_2^2}{d_2}~\frac{d_1}{\chi_2^2}\,\right)\,\right]\;f_{\rm pow}\;\,,
 \label{conf.eq}
 \ee
 where $\chi_1^2$ and $\chi_2^2$ are the value of $\chi^2$ before and after the orbital fit, respectively; while $d_1$ and $d_2$ are the
 corresponding numbers of degrees of freedom. The $f_{\rm pow}$ multiplier is required to take into account the fact that we are testing
 multiple periods $p_i$ and are selecting the one that delivers the greatest reduction in $\chi^2$. If the probability of a single
 realisation of $F$ to be less than a certain limit $F_{\rm lim}$ is $F_{\rm CDF}(F_{\rm lim})$, then the probability of the largest $F$
 among $n$ independent trials to be less than $F_{\rm lim}$ is $F_{\rm CDF}(F_{\rm lim})^n$. Thus, the $f_{\rm pow}$ multiplier in
 Equation \ref{conf.eq} is the number of {\it independent} frequencies for a given periodogram search window. For irregularly spaced
 observation times and search intervals that often exceed the Nyquist boundaries, evaluation of this number is nontrivial. We employ the
 following estimate, which is analogous to the Nyquist limit: $\,f_{\rm pow}={\rm ceil}(\Delta T/2/s_{\rm mean})\,$,
 with $s_{\rm mean}$ being the mean separation between the RV data points.

 \begin{figure}[htbp]
 \centering
 \includegraphics[angle=0,width=0.95\textwidth]{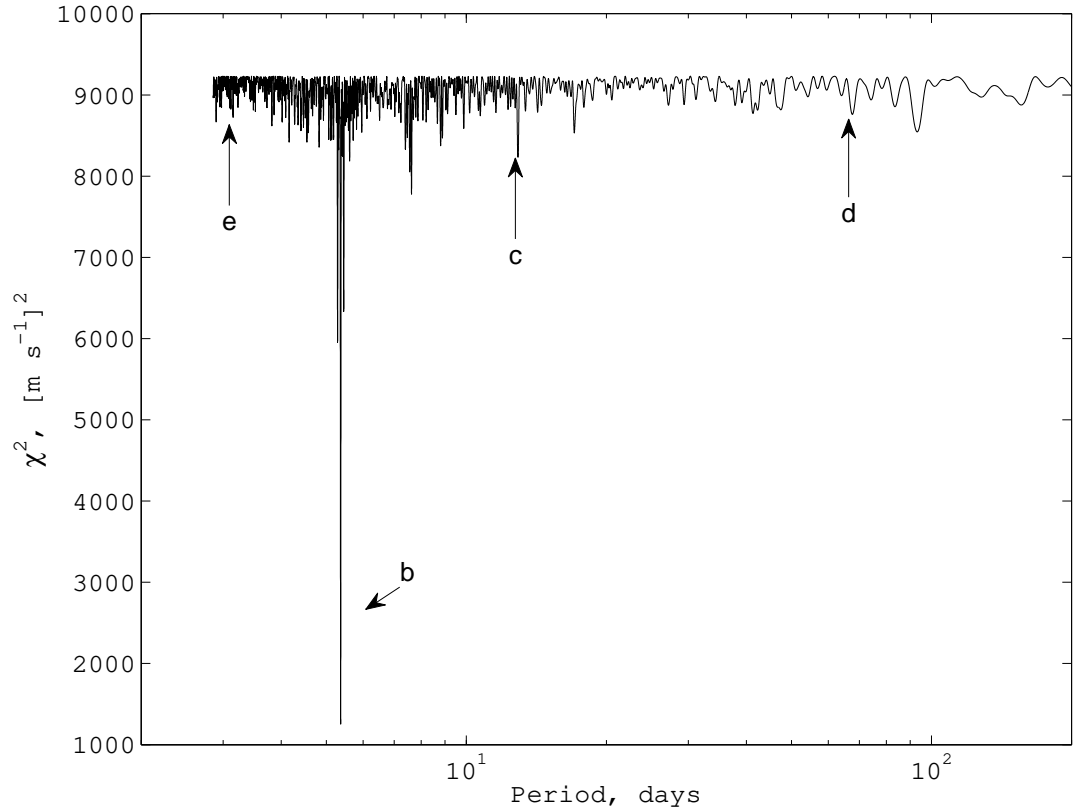}
 \caption{Generalised $\chi^2$ periodogram of the GJ 581 system prior to orbital fitting based on the HARPS data \citep{may}.
 The dips corresponding to the four detectable planets are indicated with arrows and planet names. \label{per.fig}}
 \end{figure}

 Handling of multiple planet detections is another important feature of the algorithm. Once a planet with period $P_m$ is detected, the
 corresponding terms $\,\left[\,\sum_lp_l\cos(2\pi (t_j-t_{0m})/P_m)\right.\,,$ $\left.\sum_lq_l\sin(2\pi (t_j-t_{0m})/P_m)\,\right]\,$
 get permanently added to the model. Then the search for another planet is performed outside the already detected frequencies, on the
 original RV data (which are never altered in this algorithm). With $m$ planets already detected, the size of the model is $2m+1$. Also,
 the number of degrees of freedom $d_1$ is reduced by 5 with each detected planet. The coefficients $C$ and $H$ and the RV offset are
 re-adjusted each time a new planet is detected, which provides for a better decoupling of planets in complex systems, especially of
 planets on commensurable or resonant orbits. Figure~\ref{per.fig} shows the initial $\chi^2$ periodogram of the HARPS data prior to any
 planet detection. The dips corresponding to the planets detected from these data are marked with arrows and planet designations. The
 strongest signal is associated with planet b, which shows up first. The planets are detected in their historical order: b, c, d, and e.
 The next strongest signal in the cleaned periodogram corresponds to a period of 391 d. Formally, though, it should be rejected because
 the confidence level is only 0.975.

\begin{deluxetable}{rrrrrrr}
\tablecaption{Orbital parameters of the four-planet GJ 581 system (b through e). \label{data.tab}}
\tablewidth{0pt}
\tablehead{
\multicolumn{1}{c}{Planet}  & \multicolumn{1}{c}{$P$} & \multicolumn{1}{c}{Mass} & \multicolumn{1}{c}{$a$} & \multicolumn{1}{c}{$e$} & \multicolumn{1}{c}{$\omega$} & \multicolumn{1}{c}{${\cal{M}}_0$}\\
\multicolumn{1}{c}{}  & \multicolumn{1}{c}{(d)} & \multicolumn{1}{c}{($M_\sun$)} & \multicolumn{1}{c}{(AU)} & \multicolumn{1}{c}{} & \multicolumn{1}{c}{($\degr$)} & \multicolumn{1}{c}{($\degr$)}}
\startdata
\dotfill $b$ & $5.37$  & $4.8\times 10^{-5}$ & $0.041$ & 0.01 & 288.2 & 52\\
\dotfill $c$ & $12.91$ & $1.8\times 10^{-5}$ & $0.073$ & 0.09 & 349.7 & 160\\
\dotfill $d$ & $66.98$ & $1.8\times 10^{-5}$ & $0.218$ & 0.27 & 180.9 & 92\\
\dotfill $e$ & $3.15$  & $5.5\times 10^{-6}$ & $0.028$ & 0.13 & 327.2 & 356\\
\enddata
\end{deluxetable}

The orbital and physical parameters estimated from the fits of the four detected planets are given
in Table~\ref{data.tab}. Our periods are in close agreement with the estimates by \citet{may},
but $e$ and $\omega$ (the longitude of the periastron) are markedly different. The likely reason
is that these parameters for the planets b and e were ``fixed" to 0 in \citet{may}, which is
not advisable for multiple systems. In particular, ignoring the detectable eccentricity in a least-squares adjustment
leaves in the periodogram side lobes and harmonics associated with the already detected planets
and can distort the results for other planets. Our estimate of $e$ for planet d is $0.27$,
which is lower than the $0.38\pm0.09$ obtained in {\it{Ibid}}. Our results confirm that the planets
c and e may be in the 4:1 orbital resonance, and the closeness of our estimated $\omega$
lends more credence to this possibility.

Our results are in general agreement with the work by \citet{tuo} who re-analysed the HARPS and
HIRES data together using a Bayesian approach and found evidence of only four planets existing. However,
their estimated eccentricities are consistent with 0 for all four planets. We too applied our benchmark
algorithm to the combination of HARPS and HIRES data, introducing separate RV offsets for the two data
sets. This resulted in detection of only two planets, b and c, but with significantly weaker signals.
The planet d was rejected due to insufficient confidence. Finally, the algorithm failed completely to
reproduce the results obtained by \citet{vog} from their data. {In their most recent update on the GJ 581
system, \citet{vog12} used the yet unpublished, much expanded set of RV data from the HARPS \citep{for}.
They found that a fifth planet (f) can be detected if the eccentricity of the 66.7 day planet (d) is set to
zero. Using the same data set, but not rejecting any data points in it, we were able to confirm this
statement with our detection algorithm by forcing the eccentricity of the four planets b--e to zero.
A 32.1 day planet with a projected mass of $1.99M_E$ emerges at a confidence level above 99\%. It is likely
therefore that we are dealing with the second harmonic in the signal of an eccentric planet, which can be
confused with a separate nearly commensurate planet. A potentially good way of resolving this controversy
is to look for the third harmonic of the 67 day signal, which should be generated by an eccentric
planet with that period. Indeed, we find a dip in the $\chi^2$ periodogram at $p_j\approx22.5$ days after
fitting out the four planets at zero eccentricity, but its presence can not be taken as conclusive evidence.
At $e=0.27$, the amplitudes of the first three harmonics scale as 1:0.256:0.073, hence, the expected
amplitude of the third harmonic is only $2\times0.073\simeq0.15$ m s$^{-1}$, so it should drown in the
observational noise. \citet{vog12} make a strong argument that a system of four eccentric planets can be
dynamically unstable. We therefore set out to establish that the original four-planet system as determined
from the HARPS measurements is dynamically viable.}

\section{Chaos and stability of the orbits}
\label{cha.sec}

 The preceding studies of the dynamical status and evolution of the planetary system GJ 581 have been mainly concerned with verification
 of its physical stability. \citet{beu} integrated the system of three planets (b, c, and d) for $10^8$ yr at different inclinations to
 the line of sight (which changes the estimated planetary masses by a factor of $\sin^{-1}i$) and for different orbital eccentricities.
 \citet{may} performed similar integrations for the four-planet system (b through e) with an updated period of planet d. In all these
 simulations, the orbits were assumed to be coplanar. The systems inferred from the radial velocity data proved stable for inclinations
 down to $i\simeq 30\degr$. At smaller inclinations, the masses of the super-Earth planets would become so large that the inner planet GJ
 581e would have gotten ejected quickly. \citet{beu} also computed the maximum Lyapunov exponents (MLE) and concluded that planet d is
 less chaotic than the inner planets. Within a wide range of initial parameters, the system is chaotic but long-term stable, which
 indicates that the regions of bounded chaos \citep{las90} are extensive.

  \begin{figure}[htbp]
  \centering
  \includegraphics[angle=0,width=0.95\textwidth]{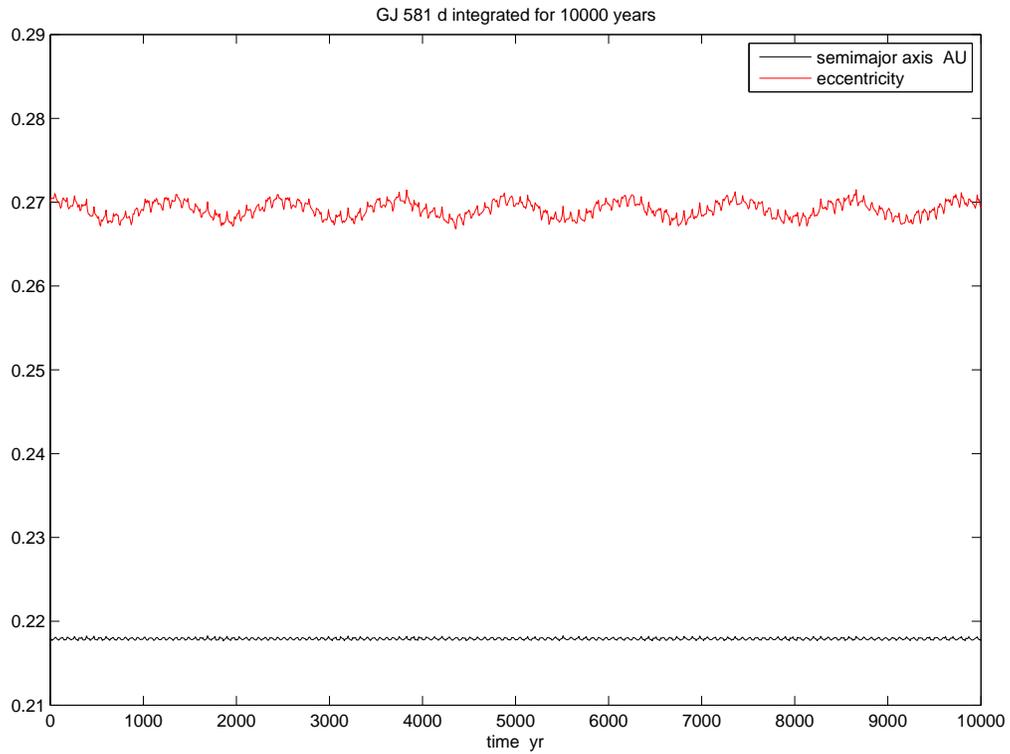}
  \caption{Orbital motion of the planet GJ 581d was integrated over $10^8$ yr forward. This figure shows the first 10,000 years of
  evolution of this planet's eccentricity (the upper line) and semimajor axis (the lower line).
  \label{ecc.fig}}
  \end{figure}
 Using the values listed in Table \ref{data.tab} as initial parameters, and assuming the orbits to be coplanar, we performed multiple
 simulations of the dynamical evolution of the five-body GJ 581 system. The symplectic integrator HNBody, version 1.0.7, \citep{rauch} was
 utilised with the symplectic option.
 We chose a time step of  $5\times 10^{-5}$ yr (0.018 days) and integrated the trajectories of all five bodies over 100 million years.
 As previous studies have found, the system appears to be quite stable. The eccentricities and semi-major axes show small periodic
 variations over the integration time. For GJ 581d, these variations over the first 10,000 years of integration are presented in Figure
 \ref{ecc.fig}. The eccentricity of this planet oscillates, seemingly forever, within a narrow range around 0.27. The semimajor axis,
 too, varies very little.

 \begin{figure}[htbp]
 \centering
 \includegraphics[angle=0,width=0.95\textwidth]{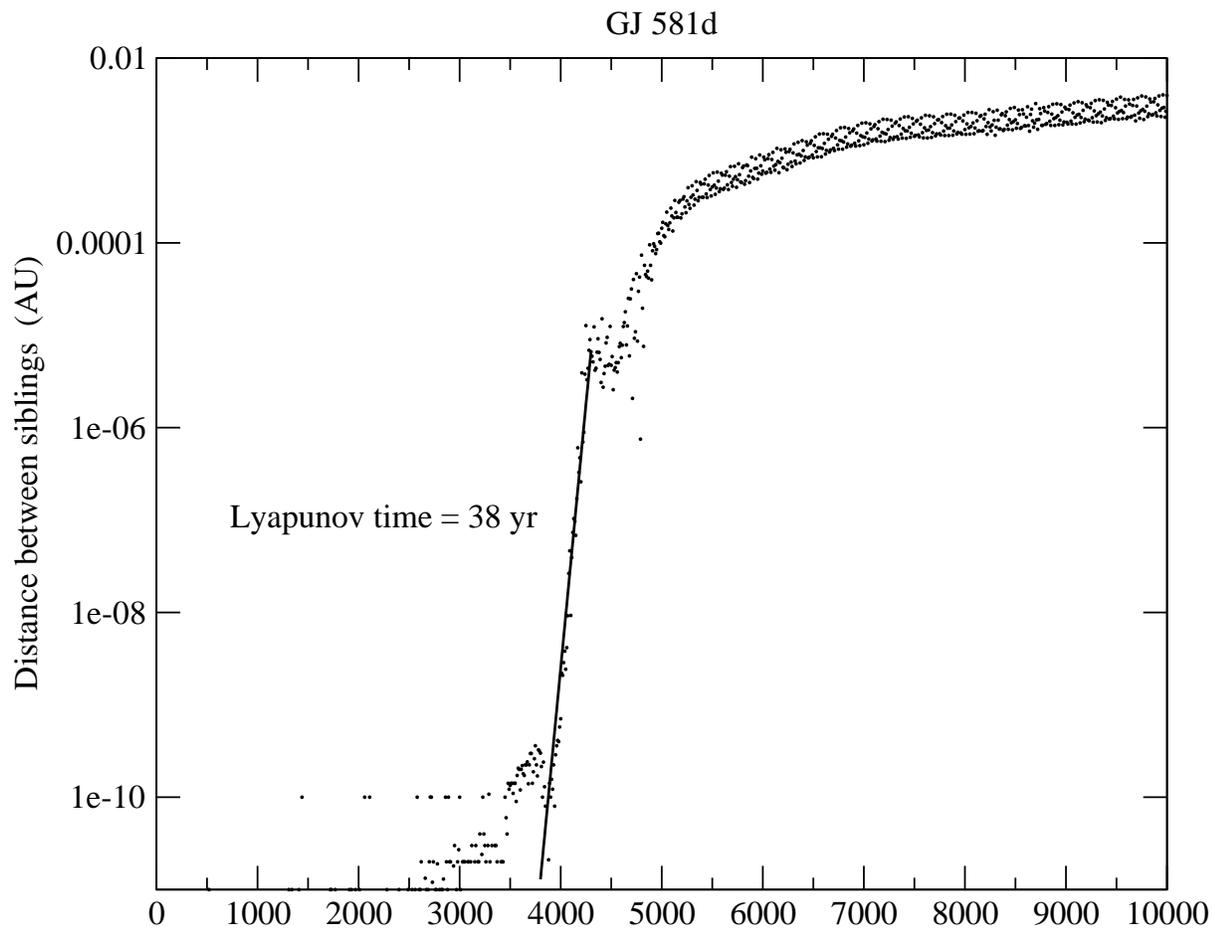}
 \caption{Distance between two sibling trajectories of the planet GJ~581d, with an initial perturbation factor in the semimajor axis
 of $10^{-14}$, integrated over 10,000 yr. The separation grows exponentially just before 4000~yr, confirming the presence of chaos in
 the orbital motion of the system. \label{sibl.fig}}
 \end{figure}

 In order to quantify the chaos in GJ 581, we employed the sibling simulation technique developed by \citet{hay1,hay2} to investigate the
 behaviour of the outer solar system with slightly different initial conditions. Two sibling trajectories are generated by perturbing the
 initial semimajor axis of the planet GJ 581b by a factor of $10^{-14}$. The distance between the unperturbed planets and their siblings
 is then computed as a function of time. Exponential divergence between the trajectories indicates that the system is chaotic. We find
 that chaos sets in within 5000 yr, with a characteristic Lyapunov time of only $\sim$ 30 yr. The distance between siblings is shown for
 planet GJ 581d in Figure \ref{sibl.fig}. In this case, the epoch of exponential divergence is close to 4000 yr, with a characteristic
 Lyapunov time of only 38 yr. The dominating planet b, on the other hand, displays two epochs of exponential divergence, with Lyapunov
 times of 18 and 277 yr.

 Long-term dynamical stability coupled with strong but bounded chaos is not uncommon among the currently known systems of multiple
 exoplanets. For example, a similar chaotic behavior was found for the dynamically robust system 55 Cnc, where a hidden commensurability
 may exist for the dominating 14 day planet. This makes the solar system, whose chaos is much slower, to stand out. In the context of our
 study, the important conclusion is that the parameters of GJ 581d listed in Table \ref{para.tab} can be safely taken constant over
 extended intervals of time.

\begin{deluxetable}{lr}
\tablecaption{Default parameters of GJ 581 and planet d.\label{para.tab}}
\tablewidth{0pt}
\tablehead{
\multicolumn{1}{c}{Parameter}  &
\multicolumn{1}{c}{Value}\\
}
\startdata
$\xi$ & \dotfill $\frac{2}{5}$ \\[1ex]
$R$ & \dotfill $1.7 R_{\rm Earth}$ \\
$M_{planet}$ & \dotfill $7.1 M_{\rm Earth}$ \\
$M_{star}$ & \dotfill $0.31 M_{\sun}$ \\
$a$ & \dotfill  $0.218$ AU\\
$e$ & \dotfill 0.27 \\
$(B-A)/C$ & \dotfill $5\times 10^{-5}$ \\
$P_{\rm orb}$ & \dotfill $67$ days \\
$\tau_M$ & \dotfill 50 yr \\
$\mu$ & \dotfill $0.8\times10^{11}$ kg m$^{-1}$ s$^{-2}$\\
$\alpha$ & \dotfill $0.2$ \\
\enddata
\end{deluxetable}

 \section{Rotational evolution of GJ 581d. A combined effect of the triaxiality and tides}
 \label{tid.sec}

 A planet is subject to the torque $\,\stackrel{\rightarrow}{{\,\cal{T}}}^{\rm{_{\,(TRI)}}}\,$ due to the planet's permanent
 triaxiality, and to the torque $\,\stackrel{\rightarrow}{{\cal{T}}}^{\rm{_{\,(TIDE)}}}\,$ generated by tidal deformation. In our
 model, we shall assume that both these torques are exerted upon the planet by its host star only. We shall thus neglect the
 torques exerted upon the planet by its moons or by other planets.

 \subsection{The equation of motion}
 \label{prec}

 Consider a planet of the mean radius $\,R\,$ and mass $\,M_{planet}\,$ and treat it as the primary. The tide-raising perturber
 (the star of mass $\,M_{star}\,$) will be regarded as the secondary effectively orbiting the primary.

 The principal moments of inertia of the planet will be denoted as $\,A,\,B,\,C\,$, in assumption that $\,A<B<C\,$. The maximal
 moment of inertia will read as $\,C=\xi M_{planet} R^2\,$, the numerical coefficient $\,\xi\,$ assuming the value of 2/5 in the
 homogeneous-sphere limit. These and other notations used in our study are listed in Table 3.

 Suppose that the planet is rotating about its major-inertia axis $\,z\,$, the one corresponding to the maximal moment of inertia
 $\,C\,$. Let $\,\theta\,$ be the sidereal angle of this rotation, reckoned from an arbitrary line fixed in inertial space (say,
 the line of apsides) to the axis $\,x\,$ of the largest elongation, i.e., the axis corresponding to the minimal moment of inertia
 $\,A\,$ of the planet.
 Rotation will then be described by the equation
 \eb
 \ddot \theta~=~\frac{{\cal{T}}^{\rm{^{\,(TRI)}}}_z+{\cal{T}}^{\rm{^{\,(TIDE)}}}_z}{C}=\frac{{\cal{T}}^{\rm{^{\,(TRI)}}}_z+
 {\cal{T}}^{\rm{^{\,(TIDE)}}}_z}{\xi M_{planet}\,R^2}\,~,
 \label{eq.eq}
 \ee
 with the subscript $\,${\it{\small{z}}}$\,$ serving to denote the polar components of the torques. High-accuracy integration of Equation
 (\ref{eq.eq}) requires a step much smaller than the orbital period.

 \subsection{The triaxiality-caused torque}

 We approximate the polar component of the torque with its quadrupole part given by \footnote{~See, e.g.,
 \citet{danb}.}
 \begin{subequations}
 \ba
 {\cal{T}}^{\rm{^{\,(TRI)}}}_z&=&\frac{3}{2}~(B-A)~\frac{{G}\,M_{star}}{r^3}~\sin2\psi\,
 \label{tri.eq_1}\\
 \nonumber\\
 &\approx&-~\frac{3}{2}~(B-A)~n^2~\frac{a^3}{r^3}~\sin2(\theta-f)\,~.
 \label{tri.eq_2}
 \ea
 \label{tri.eq}
 \end{subequations}
 Here $\,G\,$ stands for the Newton gravitational constant,  $\,r\,$ denotes the distance between the centres of mass of the two bodies, while $\,M_{star}\,$ stands for the mass of the
 star (which, in our setting, is playing the role of perturbing secondary effectively orbiting the tidally deformed primary). The
 mean motion is given by $~n\equiv\sqrt{G(M_{planet}+M_{star})/a^{3}\,}\,\approx\,\sqrt{GM_{star}/a^3\,}~$, in understanding that
 the star is much more massive than the planet.

\begin{figure}[htbp]
 \centering
 \includegraphics[
 width=0.95\textwidth
 ]{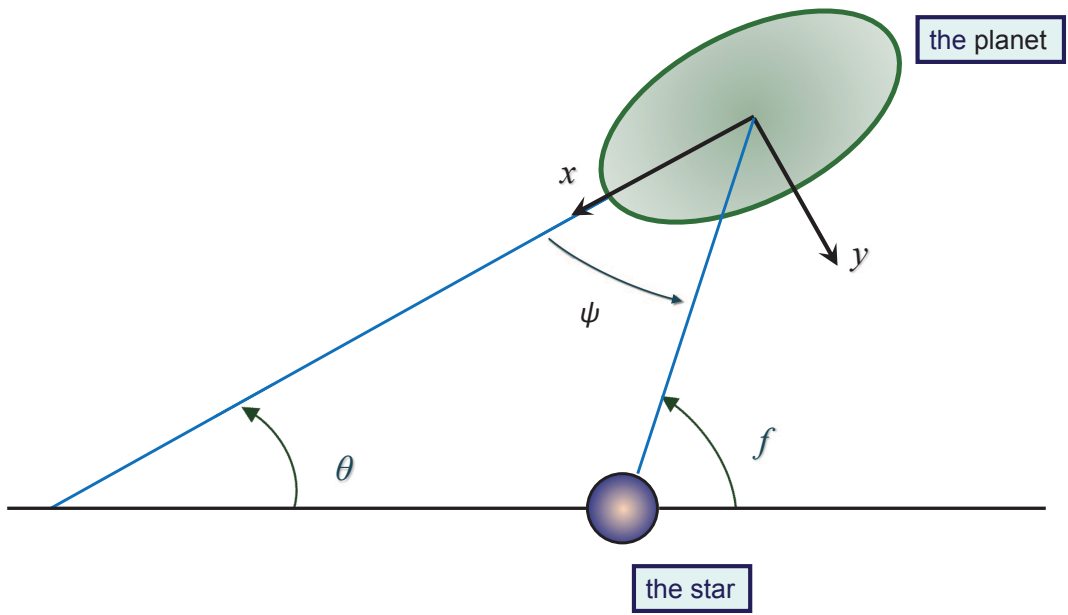}
 \caption{Horizontal line is that of apsides, so $\,f\,$ is the true anomaly, $\,\theta\,$ is the sidereal angle of
 the planet, while $\,\psi\,=\,f\,-\,\theta\,$. The principal axes $\,x\,$ and $\,y\,$ of the planet correspond to the minimal and middle moments of inertia, appropriately.
 \label{Figure}}
 \end{figure}

 Our notations for angles are given on Figure \ref{Figure}.
 The angle $\,\psi\,$ renders separation between the planetocentric
 direction towards the star and the principal axis $\,x\,$ of the maximal elongation (the minimal-inertia axis). It is
 equal to the difference between the angles $\,f\,$ and $\,\theta\,$ made by these two directions on an arbitrary line fixed in inertial
 space. If this fiducial line is chosen to be parallel to the line of apsides connecting the star with the empty focus, then
 $\,f\,$ will be the planet's true anomaly as seen from the star, while $\,\theta\,$ will be the sidereal angle (which we agreed
 to reckon from the line of apsides.).

 Employing (\ref{tri.eq_2}) in practical computations, it is advantageous to approximate the functions $\,(a/r)^3\sin2f\,$
 and $\,(a/r)^3\cos2f\,$ with truncated series of the Hansen's coefficients \citep[e.g.,][]{bra}, which in their turn can be
 approximated with series of the Bessel functions of the first kind.

 \subsection{The tidal torque}\label{tt}

 Due to a several-orders-of-magnitude difference in the intensities of internal friction in a terrestrial planet and a star, we shall
 take into account only the tides exerted by the star on the planet, and shall neglect the tides on the star.

 The technique of calculation of tidal torques had until recently remained in a somewhat embryonic state requiring much of correction.
 On the one hand, it had long been habitual and common to combine the expression for the tidal torque with unphysical rheological models.
 As a result, much of the hitherto obtained results on the spin history of terrestrial planets \citep[e.g.,][]{cel,hel} have to be
 reexamined, because they were based on very {\it{ad hoc}} rheologies incompatible with the behaviour of realistic solids. On the other
 hand, in many publications the description of the tidal torque was marred by a mathematical mistake which proliferated through many
 papers and ended up in textbooks. Explanation and correction of that long-standing oversight is presented in \citet{e&m}.

 We shall limit our treatment to the simpler case where the planet is not too close to the star (${\textstyle R}/{\textstyle a}\ll1$),
 the obliquity of the planet is small ($i\simeq 0$), and its eccentricity $\,e\,$ is not very large. As demonstrated in the Appendix A,
 under these assumptions the polar component of the secular part of the tidal torque can be approximated by
 \ba
 \nonumber
 \langle\,{\cal{T}}_z^{\rm{_{\,(TIDE)}}}\rangle_{\textstyle{_{\textstyle_{\textstyle{_{l=2}}}}}}~=~~\quad~\quad~~\quad~\quad~\quad~\quad~
 \quad~\quad~\quad~\quad~\quad~\quad~\quad~\quad~\quad~\quad~\quad~\quad~\quad~\quad~\quad~\quad~\quad~\quad~\quad~\quad\\
 \nonumber\\
 \frac{3}{2}~G\,M_{star}^{\,2}\,R^5\,a^{-6}\sum_{q=-1}^{7}\,
 G^{\,2}_{\textstyle{_{\textstyle{_{20\mbox{\it{q}}}}}}}(e)~k_2(\omega_{\textstyle{_{\textstyle{_{220\mbox{\it{q}}}}}}})~
 \sin|\,\epsilon_2(\omega_{\textstyle{_{\textstyle{_{220\mbox{\it{q}}}}}}})\,|
 \,~\mbox{Sgn}\,\left(\,\omega_{220q}\,\right)
 +O(e^8\,\epsilon)+O(\inc^2\,\epsilon)
 ~~,~\quad~\quad~
 \label{16a}
 \label{tid.eq}
 \ea
 where $\,R\,$ denotes the radius of the planet, $\,\dot{\theta}\,$ stands for its rotation rate, while $\,a\,$, $\,e\,$, and $\,i\,$
 are the semimajor axis, the eccentricity, and the obliquity of the perturber (the star) as seen from the planet. The notation
 $\,G_{lpq}(e)\,$ stands for the so-called eccentricity polynomials which coincide with the Hansen polynomials
 $\,X^{\textstyle{^{(-l-1),\,(l-2p)}}}_{\textstyle{_{(l-2p+q)}}}(e)~$.

 For the first time, this expression appeared, without proof, in the work by \citet{gold} who summed over all integer values of $\,q\,$. For the reasons explained
 in Appendix A below, we limit the summation to the span from $\,q=-1\,$ through $\,q=7\,$. This truncation effectively limits the Taylor expansions of
 eccentricity-dependent functions to terms of the orders up to $\,e^7\,$, inclusive. The resulting relative precision of our calculations is approximately
 0.1\%$\,$, because the largest and the smallest terms of the so-truncated sum contain $\,G_{200}(0.27)=0.82202\,$ and $\,G_{207}(0.27)=0.01392\,$, respectively; while all the eccentricity polynomials $\,G_{20q}(e)\,$ outside the range of summation (i.e., for $\,q\,$ outside of the range $~-\,1,\,.\,.\,.\,,\,7\,$)  assume much smaller values.

 In Equation (\ref{16a}), both the dynamical Love number $\,k_2\,$ and the phase lag $\,\epsilon_2\,$ are functions of the tidal Fourier
 mode which in this, simplified case reads as
 \ba
 \omega_{\textstyle{_{220\mbox{\it{q}}}}}\,=\,(2+q)\;n\,-\,2\;\dot{\theta}\,~,~~~
 \label{LL9}
 \ea
 a more general expression for $\,\omega_{lmpq}\,$ being given in the Appendix A. As ever, $\,n\,$ denotes the mean motion.

 The dynamical Love numbers $\,k_2(\omega_{\textstyle{_{220\mbox{\it{q}}}}})\,$ are positive definite, while the sign of each lag
 $\,\epsilon_2(\omega_{\textstyle{_{220\mbox{\it{q}}}}})\,$ coincides with that of the tidal mode $\,\omega_{\textstyle{_{220\mbox{\it{q}
 }}}}\,$, as explained in Appendix A. Therefore each factor $\,k_2(\omega_{\textstyle{_{220\mbox{\it{q}}}}})\,\sin\epsilon_2
 (\omega_{\textstyle{_{220\mbox{\it{q}}}}})\,$ can be written as $\,k_2(\omega_{\textstyle{_{220\mbox{\it{q}}}}})\,\sin|\,\epsilon_2
 (\omega_{\textstyle{_{220\mbox{\it{q}}}}})\,|~\,\mbox{Sgn}\,(\omega_{\textstyle{_{220\mbox{\it{q}}}}})\,$.

 Through Equation (\ref{LL9}), each such factor becomes a function of the planetary spin rate $\,\dot{\theta}~$:
 \ba
 \nonumber
 k_2(\omega_{\textstyle{_{220\mbox{\it{q}}}}})\;\sin|\,\epsilon_2(\omega_{\textstyle{_{220\mbox{\it{q}}}}})\,|~\,\mbox{Sgn}\,(\,\omega_{
 \textstyle{_{220\mbox{\it{q}}}}}\,)~\quad\quad\quad\quad\quad\quad\quad\quad~\quad\quad\quad\quad\quad\quad\quad\quad\quad\quad\quad\quad
 ~\\  \nonumber\\
 =~k_2(\,2(n-\dot{\theta})\,+\,q\,n\,)~~\sin|\,\epsilon_2(\,2(n-\dot{\theta})\,+\,q\,n\,)\,|~~\mbox{Sgn}\,(\,2(n-\dot{\theta})
 \,+\,q\,n\,)\,~.~\quad
 \label{formula}
 \ea
 Accordingly, the entire sum (\ref{16a}) can be regarded as a function of $\,\dot{\theta}\,$. The mean motion and eccentricity act as
 parameters, their evolution being much slower than that of $\,\dot{\theta}~$.

\begin{figure}[htbp]
 \centering
 \includegraphics[angle=0,width=0.95\textwidth]{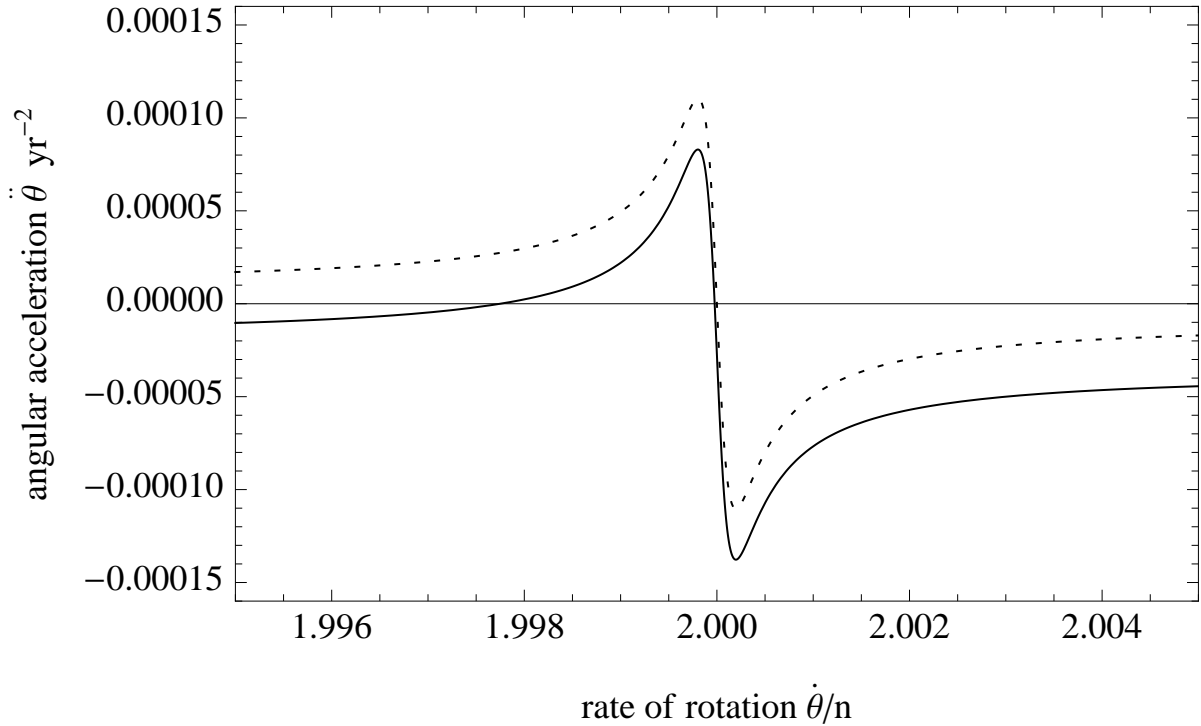}
 \caption{Angular acceleration of the planet GJ 581d caused by the secular tidal torque (\ref{16a}), in the vicinity of the 2:1 spin-orbit resonance.
 The dotted kink-shaped curve depicts the term $\,q\,=\,q\,'\,$, which is an odd function when centered at $~\dot{\theta}/n\,=\,1\,+\,q\,'/2~$.
 In the paper, we denote this term with $\,W(\dot{\gamma})\,$.
 The  solid line furnishes the overall torque calculated as a sum of the $\,q\,=\,q\,'\,$ kink and the bias. The bias, denoted in the text with $\,V\,$,
 is comprised by the terms of Equation (\ref{16a}), with $\,q\,\neq\,q\,'\,$.
 \label{tide.fig}}
 \end{figure}

 In each term of series (\ref{16a}), the factor $\,k_2\,\sin\epsilon_2\,$, expressed as a function of $\,\dot{\theta}\,$, has the shape
 of a kink. In Figure \ref{tide.fig}, the dotted line \footnote{~In a hypothetical case of a planet despinning at a constant rate through a resonance,
 the appropriate tidal mode becomes linear in time. Then the tidal torque assumes a similar kink shape, as a function of time. This situation is
 considered in Ferraz-Mello (2012, Figure 7b). Any physically reasonable rheology must lead to this or similar type of tidal torque behaviour in the
 vicinity of a resonance.} illustrates the behaviour of the factor
 \ba
 \nonumber
 k_2(\omega_{\textstyle{_{
 \textstyle{_{2202}}}}})~\sin\epsilon_2(\omega_{\textstyle{_{\textstyle{_{2202}}}}})\,=\,
 k_2(\,4\,n\,-\,2\,\dot{\theta}\,)~\sin\epsilon_2(\,4\,n\,-\,2\,\dot{\theta}\,)\\
 \nonumber\\
 =~k_2(\,4\,n\,-\,2\,\dot{\theta}\,)~~\sin|\,\epsilon_2(\,4\,n\,-\,2\,\dot{\theta}\,)\,|~~\mbox{Sgn}\,(\,4\,n\,-\,2\,\dot{\theta}\,)\,~.
 \label{}
 \ea

 The origin of the kink shape is explained in Appendix B, with references provided. From the physical viewpoint, the emergence of this
 shape is natural, in that each term should transcend zero and change its sign smoothly when the spin rate goes through the appropriate
 spin-orbit resonance.

 However, in Equation (\ref{16a}) the kink-shaped factors are accompanied with different multipliers $\,G^{\,2}_{\textstyle{_{\textstyle{_{20q}}}}}
 (e)\,$. So the sum (\ref{16a}), as a function of $\,\dot{\theta}~$, will be a superposition of many kinks having different magnitudes
 and centered at different resonances
 (nine kinks, if the sum over $\,q\,$ goes from -1 to 7). The resulting curve will cross the horizontal axis in points
 {\it{close to}} the resonances $\,\dot{\theta}\,=n\,\left(1+\,{\textstyle q}/{\textstyle 2}\right)\,$, but not exactly in these
 resonances --- see the solid line in Figure \ref{tide.fig}.

 \citet{gol68} were first to point out that in the vicinity of a particular resonance $\,q\,'\,$, i.e., for
 $\,\dot{\theta}\,$ being close to $\,n\,\left(1\,+\,{\textstyle q\,'}/{\textstyle 2}\right)\,$, the right-hand side of Equation (\ref{tid.eq})
 can be conveniently decomposed into two parts. One part is the $\,q=q\,'\,$ term,
 an odd function vanishing at $\,\dot{\theta}\,=\,n\,\left(1\,+\,{\textstyle q\,'}/{\textstyle 2}\right)\,$. As was
demonstrated later by Efroimsky (2012a,b), for realistic rheologies this function has the
shape of a kink --- see the dotted line in Figure 5.
 Another part, called {\it{bias}}, is comprised by the rest of the
 sum. So the bias is the contribution of all the $\,q\neq q\,'\,$ modes into the values assumed by the torque in the vicinity of the $\,
 q=q\,'\,$ resonance. For not too large eccentricities, the bias is usually negative in value. The bias is a very slowly changing
 function, which can, to a good approximation, be treated as constant.

 The $\,q=q\,'\,$ term by itself is an odd function, and it goes through nil at exactly $\,\dot{\theta}\,=\,n\,\left(1\,+\,{\textstyle q\,'}/{\textstyle 2}\right)\,$. However, the bias displaces the location of zeroes. In Figure \ref{tide.fig}, the torque (depicted with a solid line) is defined mostly by the term with $\,q=2\,$ (rendered by the dotted line). However, the curve is shifted down due to the bias which is defined mainly by the right slope of the $\,q=1\,$ kink located to the left. As the right slope of the $\,q=1\,$ kink is negative, the $\,q=2\,$ kink in Figure \ref{tide.fig} is shifted slightly down, and the zero is located slightly to the left of the point $\,\dot{\theta}\,=\,2\,n\,$. The crossing point in this case is at $\,\dot\theta/n=1.999976481\,$. Such a minuscule shift of the equilibrium away from the resonance frequency does not have any practical consequences, because the net nonzero tidal torque is compensated by a tiny secular triaxial torque, as discussed in more detail in \citet{m&e}.

 The shifts of the tidal-equilibrium frequencies at resonances are larger for bodies with a lower Maxwell time, e.g., for bodies whose mantles contain a large fraction of partial melt. Still, no matter how shifted the tidal equilibrium happens to be, the mean rotation is exactly resonant due to the presence of the compensating triaxiality-caused secular torque.

 \begin{deluxetable}{lr}
 \tablecaption{Symbol key \label{nota.tab}}
 \tablewidth{0pt}
 \tablehead{
 \multicolumn{1}{c}{Notation}  &
 \multicolumn{1}{c}{Description}\\
 }
 \startdata
 $\xi$ & \dotfill moment of inertia coefficient of the planet\\
 $R$ & \dotfill radius of the planet \\
 ${\cal{T}}^{\rm{^{\,(TRI)}}}$ & \dotfill triaxiality-caused torque acting on the planet\\
  ${\cal{T}}^{\rm{^{\,(TIDE)}}}$ & \dotfill tidal torque  acting on the planet\\
 $M_{planet}$ & \dotfill mass of the planet \\
 $M_{star}$ & \dotfill mass of the star \\
 $a$ & \dotfill semimajor axis of the planet \\
 $r$ & \dotfill $.\,.\,.\,$instantaneous distance between the planet and the star \\
 $f$ & \dotfill true anomaly of the planet \\
 $e$ & \dotfill orbital eccentricity \\
 ${\cal{M}}$ & \dotfill mean anomaly of the planet \\
 $C$ & \dotfill the maximal moment of inertia of the planet \\
 $B$ & \dotfill the middle moment of inertia of the planet \\
 $A$ & \dotfill the minimal moment of inertia of the planet \\
 $n$ & \dotfill mean motion, i.e. $\,\sqrt{G(M_{planet}+M_{star})/a^{3}\,}\,$
 ~\\
 $G$ & \dotfill gravitational constant, $=66468$ m$^3$ kg$^{-1}$ yr$^{-2}$ \\
 $\tau_{_M}$ & \dotfill viscoelastic characteristic time (Maxwell time) \\
 $\tau_{_A}$ & \dotfill inelastic characteristic time (Andrade time)\\
 $\mu$ & \dotfill unrelaxed rigidity modulus \\
 $J$ & \dotfill unrelaxed compliance \\
 $\alpha$ & \dotfill the Andrade parameter \\
 \enddata
 \end{deluxetable}

 \section{Probabilities of capture into resonances. General facts}

 When the planet's spin rate approaches a resonance $2\,\dot\theta=(2+q)n\,$, with an integer $q$, the planet can be captured, or can
 traverse the resonance, depending on the specific trajectory in the phase space. A method for estimation of capture probabilities was
 developed by \citet{gold}, for two simplified models of the tidal torque. A clearer explanation of this theory was presented in the
 later work by \citet{gol68}, to which we shall refer.

 In one model considered in these papers, the torque was assumed to be linear in the tidal frequency. To be more precise, in the Fourier
 expansion of the torque over tidal modes, each torque component was set to be proportional to the appropriate tidal mode \footnote{~The
 term {\it{tidal mode}} is more appropriate than {\it{frequency}}, because a Fourier mode can assume either sign, while the physical
 frequencies are the modes' absolute values and thus are positive definite.} -- see formula (19) in \citet{gol68}.

 Another model addressed in the said two works was the constant-torque one. Specifically, in the Fourier expansion of the torque over
 tidal modes, each torque component was set to be a constant multiplied by the sign of the corresponding mode -- see Equation (29) in
 \citet{gol68}.

 In their treatment of both models, \citet{gol68} took into account only the secular, orbit-averaged, component of the tidal
 torque, and ignored the existence of an oscillating component. Below we shall test the validity of this approximation.

 The pivotal ideas and formulae of the capture theory are explained in short in Appendix C below. Here we shall employ some of
 those formulae, though with an important difference. Instead of the toy models introduced in \citet{gol68}, we shall rely on a
 realistic rheology of solids. As we shall see, capture probabilities are sensitive to changes in (at least some of the) rheological
 parameters. For example, the probabilities turn out to be more sensitive to the mantle's Maxwell time than to the planet's triaxiality.

 The principal result of the analysis carried out by \citet{gol68} is their estimate for the probability of capture into an
 arbitrary resonance $\,q\,=\,q\,'\,$, i.e., into a steady rotation at the rate of $\,\dot{\theta}\,=\,n\,\left(1\,+\,{\textstyle q\,'}/
 {\textstyle 2}\right)\,$. The probability is given by
 \eb
 P_{\rm capt}\,=\,\frac{\textstyle 2}{\textstyle{1\,+\,{\textstyle 2~\pi~V}/{\textstyle \int_{-\pi}^{\pi}\,W(\dot\gamma)\,d\gamma}}}\,~,
 \label{probability}
 \ee
 where the new, auxiliary variable $\,\gamma\,$ is defined through
 \ba
 \gamma~\equiv~2~\theta~-~(2\,+\,q\,')\,{\cal{M}}\,~,
 \label{}
 \ea
 its time derivative thus being the negative double of the tidal mode $\,\omega_{\textstyle{_{220\mbox{\it{q}}\,'}}}~$:
 \ba
 \dot\gamma\,=\,-~(2\,+\,q\,')\,n\,+\,2\,\dot{\theta}\,\,=\,-\,\omega_{220\mbox{\it{q}}\,'}\,~.
 \label{}
 \ea
 This mode vanishes in the $\,q\,=\,q\,'\,$ resonance, so we may say that this resonance corresponds to $\,\dot\gamma\,=\,0~$.
 The odd function $\,W(\dot{\gamma})\,$ entering expression (\ref{probability}) is called $\,${\it{kink}}$\,$ and is given by
 \ba
 W(\dot{\gamma})~=~-~K~G^{\,2}_{\textstyle{_{220\mbox{\it{q}}\,'}}}~k_2(\dot{\gamma})~\sin |\,\epsilon_2(\dot{\gamma})\,|~\,\mbox{Sgn}
 \,(\dot{\gamma})\,~,
 \label{WW}
 \ea
 $K\,$ being a positive constant. Be mindful that our definition of $\,\gamma\,$ is in agreement with that used by \citet{makti} and
 is twice that introduced in \citet{gold,gol68}.

 Entering formula (\ref{probability}) is also a function \footnote{~Writing the right-hand side of Equation (\ref{VV}), we used the fact
 that we are in the vicinity of the $\,q\,=\,q\,'\,$ resonance -- see formula (\ref{13}) in Appendix C.}
 \ba
 \nonumber
 V&=&K\,\sum_{q\neq q\,'}G^{\,2}_{220\mbox{\it{q}}}~k_2(\omega_{\textstyle{_{220\mbox{\it{q}}}}})~\sin |\,
 \epsilon_2(\omega_{\textstyle{_{220\mbox{\it{q}}}}})\,|~\,\mbox{Sgn}\,(\omega_{\textstyle{_{220\mbox{\it{q}}}}})\\
 \nonumber\\
 &=&K\,\sum_{q\neq q\,'}G^{\,2}_{220\mbox{\it{q}}}~k_2(\,(q\,-\,q\,')\,n\,)~\sin |\,
 \epsilon_2(\,(q\,-\,q\,')\,n\,)\,|~\,\mbox{Sgn}\,(q\,-\,q\,')\,~.
 \label{VV}
 \ea
 called $\,${\it{bias}}$\,$. The physical meaning of $\,V\,$ and $\,W(\dot{\gamma})\,$ is explained in Appendix C. The Appendix also
 explains the way how in the above integral the dependence of $\,\dot{\gamma}\,$ upon $\,\gamma\,$ should be set, so that the integral
 could be evaluated.

 \begin{figure}[htbp]
 \centering
 \plottwo{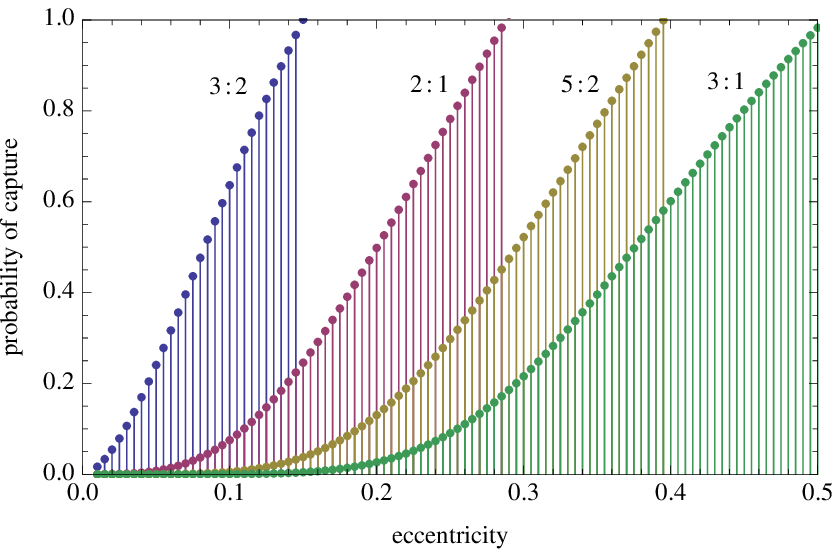}{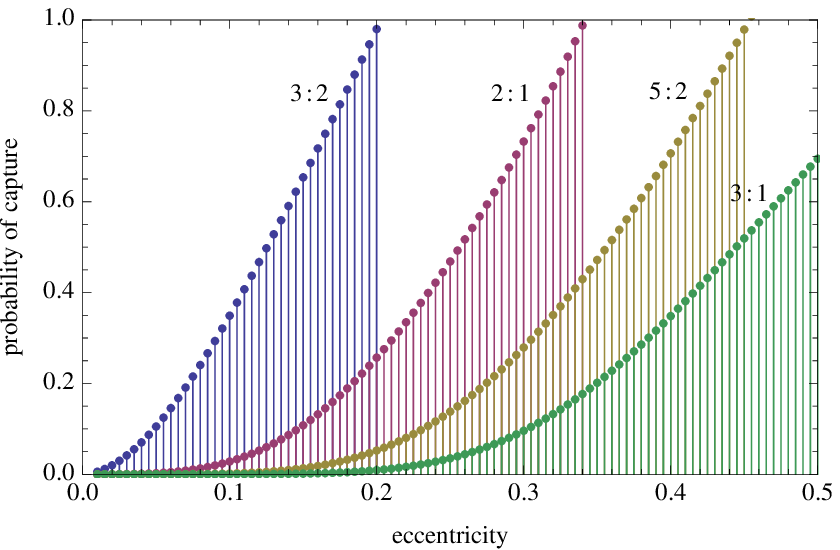}
 \caption{Probabilities of capture of GJ581d at 3:2, 2:1, 5:2, and 3:1 spin-orbit resonances as functions of eccentricity,
 for $\,(B-A)/C=5\times 10^{-5}\,$. $\,$Left: a warmer planet, with a lower viscosity and $\,\tau_M=50$ yr. $\,$Right:
 a colder planet, with a higher viscosity and $\,\tau_M=500$ yr.
 \label{prob.fig}}
 \end{figure}

  \section{Probabilities of capture into resonances. The case of a realistic rheology}

 Important to us is the fact that the above expressions for $\,V\,$ and $\,W(\dot{\gamma})\,$ contain the mode-dependent factor $\,k_2
 (\omega_{\textstyle{_{220\mbox{\it{q}}}}})\,\sin\epsilon_2(\omega_{\textstyle{_{220\mbox{\it{q}}}}})\,$. The functional form of its
 mode-dependence is defined by the self-gravitation and rheology of the planet \citep{efr1}. This observation helps us to mark the point
 at which our analysis will diverge from that carried out by \citet{gol68}$\,$: our choice of a realistic rheology will yield for $\,k_2
 \,\sin\epsilon_2\,$ a mode-dependence different from both cases addressed in {\it{Ibid}}.

 Recently, \citet{makti} generalised the treatment by \citet{gol68} for the case of a more physical tidal torque, i.e., for a torque whose
 terms bear a mode-dependence stemming from the properties of realistic solids. The rheology of silicates and ices is well described by
 the Andrade model, insofar as the tidal frequency is sufficiently high and the inner friction is dominated by defect unpinning. At lower
 frequencies, the friction is dominated by viscosity, and the behaviour of the material becomes closer to that of the Maxwell body
 \citep{efr1,efrb}. The resulting mode-dependencies of the terms of the torque are presented below in Appendix B. These mode-dependencies
 are more complicated than the two models considered by \citet{gol68}.

 In our computations, we ignore the oscillatory components of the tidal torque, and consider only the secular part which can be obtained by averaging over one
 orbital period --- a convenient simplification is to be justified below. Thus we begin in the spirit of \citet{gol68}, i.e., use formula (\ref{probability}),
 the functions $\,W\,$ and $\,V\,$ being furnished by formulae (\ref{WW}) and (\ref{VV}), correspondingly. Following \citet{makti}, we
 then insert into those formulae the realistic mode-dependence of $\,k_2\,\sin\epsilon_2\,$ written down and explained in Appendix B.
 Finally, we perform a brute force numerical check of whether the neglect of the oscillating part is legitimate.

 The resulting capture probabilities as functions of eccentricity are presented in Figure \ref{prob.fig} for two values of the Maxwell time: $\,\tau_{_M}=50$ yr (the
 left plot) and $\,\tau_{_M}=500$ yr (the right plot). While $\,500$ yr is the current Maxwell time of Earth's mantle, the value of $\,50$ yr is deemed to represent a slightly \footnote{~We say $\,${\it{slightly}}$\,$, because the viscosity $\eta$ depends upon the temperature $T$ through the Arrhenius formula $\,\eta \propto \exp(A^*/RT)\,$, while the rigidity $\,\mu\,$ depends upon $\,T\,$ slower, unless we are close to the melting point (the latter caveat being hardly relevant, as the lithostatic pressure prevents the mantle from melting -- even though some partial melt may be present). So the Maxwell time $\,\tau_{_M} \equiv \eta / \mu\,$ depends on the temperature about exponentially. A small variation of the temperature renders the following relative changes of the viscosity
 and the Maxwell time: $~-\,\frac{\textstyle \Delta \tau_{_M}}{\textstyle \tau_{_M}}\,\approx\,-\,\frac{\textstyle \Delta \eta}{\textstyle \eta}\,\approx~
 \frac{\textstyle \Delta T}{\textstyle T^2} ~\frac{\textstyle A^*}{\textstyle R}~$. Here the activation energy may be estimated, for olivines and silicate perovskites, as $\,A^*\approx 6 \times 10^5\,$ J mol$^{-1}$.
 Consider an Earth-like planet with a silicate mantle of a mean temperature $\,T=2300$ K. A decrease of the viscosity and the Maxwell time by 9/10 would
  correspond to an increase of the temperature by $\,\Delta T = 66$ K. For the Earth, it would
 imply a less than 1 Gyr step back to the Neoproteozoic Era, one marked with appearance of the first multicelled organisms. Therefore both values, $\,500\,$ yr and
 $\,50\,$ yr, serve as legitimate estimates for the Maxwell time of an Earth-like, potentially habitable planet.} warmer planet, one with a lower viscosity.

 From the plots, we see that probabilities of capture are sensitive to the Maxwell time. Warmer planets with lower
 $\,\tau_M\,$, have more chances of being captured at spin-orbit resonances, as they loose their spin angular momentum. In particular,
 for $\,\tau_M=50$ yr and $\,(B-A)/C\,=\,5\times 10^{-5}\,$, the capture probabilities are: 1 in resonance 3:2, 0.897 in 2:1, 0.383 in 5:2,
 0.133 in 3:1, and 0.040 in 7:2. Therefore, if GJ 581d has had enough time to de-spin into a state of spin-orbit equilibrium at the
 current value of eccentricity, then the probabilities of the planet being now in resonances are the following: 0.05 in resonance 3:2,
 0.46 in 2:1, 0.32 in 5:2, 0.13 in 3:1, and 0.04 in 7:2.

 For $\,\tau_M=500$ yr and $\,(B-A)/C=5\times 10^{-5}\,$, the capture probabilities are: 1 at resonance 3:2, 0.568 in 2:1, 0.188 in 5:2,
 0.054 in 3:1, and 0.014 in 7:2. Then, with a probability of 0.35, the planet is currently entrapped in resonance 3:2, 0.43 in 2:1, 0.18 in
 5:2, 0.05 in 3:1, and 0.01 in 7:2. The 2:1 resonance is the most likely state of the planet in both cases. However, for $\tau_M=50$ yr
 the second likeliest state is 5:2, while for $\tau_M=500$ yr it is 3:2.

 The probabilities of capture depend on the degree of triaxiality through the parameter $\,(B-A)/C\,$ in the equation of separatrix (\ref{separ.eq}). This
 equation tells us that larger values of $\,(B-A)/C\,$ entail greater libration amplitudes of $\,\dot\gamma\,$. The kink $\,W(\dot\gamma)\,$, which is the antisymmetric part of the secular torque, becomes smaller with $\,\dot\gamma\,$ growing outside the resonance, see Figure \ref{tide.fig}. Therefore, the integral $\,\int_{-\pi}^{\pi}\,W(\dot\gamma)\,d\gamma\,$ in Equation (\ref{prob.eq}) is expected to become smaller for larger $\,(B-A)/C\,$. Accordingly, the capture probability is to become
 smaller for larger $\,(B-A)/C\,$. This is confirmed by computations of capture probabilities for $\,\tau_M=50$ yr, $\,(B-A)/C=2\times 10^{-4}\,$ and the other parameters as given in Table \ref{para.tab}. The capture probabilities are: 1 in the resonance 3:2, 0.776 in 2:1, 0.301 in 5:2, 0.097 in 3:1, and 0.028 in 7:2.
 Each of these probabilities is slightly smaller than its counterpart taken for the smaller \footnote{~For terrestrial planets, as well as for large solid
 moons, our choice of the values $\,5\cdot10^{-5}\,$ and $\,2\cdot10^{-4}\,$ for $\,(B-A)/C\,$ is likely to be realistic. Recall that $\,(B-A)/C\,$ is equal
 to $\,2.2\times 10^{-5}\,$ for the Earth (Wen-Bin Shen et al. 2011, Liu \& Chao 1991), to $\,6.9\times 10^{-4}\,$ for Mars (Edvardsson et al. 2002), and
 to $\,2.3\times 10^{-4}\,$ for the Moon (Williams et al. 1996).} triaxiality $\,(B-A)/C=5\cdot10^{-5}\,$. At the same time, the relatively small
 differences between the capture probabilities into the same resonance, for different values of $\,(B-A)/C\,$, imply that the spin-orbit resonances are
 more sensitive to $\,\tau_M\,$ than to $\,(B-A)/C\,$.

 As a way of spot-check verification of these theoretical results, we conducted brute-force simulations of the GJ 581d system. The
 equation of motion, incorporating both the tidal and triaxial torques acting on the planet, was numerically integrated 40 times for
 $\,\tau_M=50$ yr, $\,(B-A)/C=5\cdot10^{-5}\,$, $\,e=0.27\,$, the initial spin rate $\,\dot\theta(0)=2.51\,n\,$, the initial mean anomaly
 $\,{\cal{M}}(0)=0\,$, and the initial sidereal angle $\,\theta(0)=\pi\,i/40\,$, with $\,i=0,1,\ldots,39\,$. This method of estimation
 tacitly assumes that the initial sidereal angle at a fixed rate of rotation can take any values with equal probability (which is less
than obvious). With this assumption accepted, simulations spanning 7000 yr, with a
step of $\,1.5\cdot10^{-3}$ yr, resulted in 14 captures into the 5:2 resonance and 26 passages. The estimated probability of capture is thus $\,14/40=0.35\,$, which is surprisingly close to the theoretical
 estimate $\,0.383\,$. To make the numerical and theoretical estimations consistent, the former included only the secular part of the
 torque (\ref{tid.eq}). This points at one of the weaknesses of the semi-analytical derivation of probabilities. The oscillatory terms of
 the tidal torque are ignored altogether. \footnote{~Recall that the common expression (\ref{16a}) for the polar component of the tidal
 torque renders only its secular part. It is for this reason, that in this formula we use the angular brackets denoting an average over
 an orbital period. A full expression for the torque includes also an oscillating part which averages out over an orbital period, but
 which may nonetheless play a role in the capture process \citep{efrb}.}$\,$ Furthermore, even though the deformity of the planet
 $\,(B-A)/C\,$ enters the computation of capture probabilities, the cyclic variations of the spin rate caused
 by the triaxial torque are not involved in any way. In reality, however, the smooth sinusoidal separatrix trajectories defined by Equation (\ref{separ.eq})
 are superimposed with a jitter caused by the harmonics of time-dependent terms of the torque.
 \begin{figure}[htbp]
 \centering
 \plotone{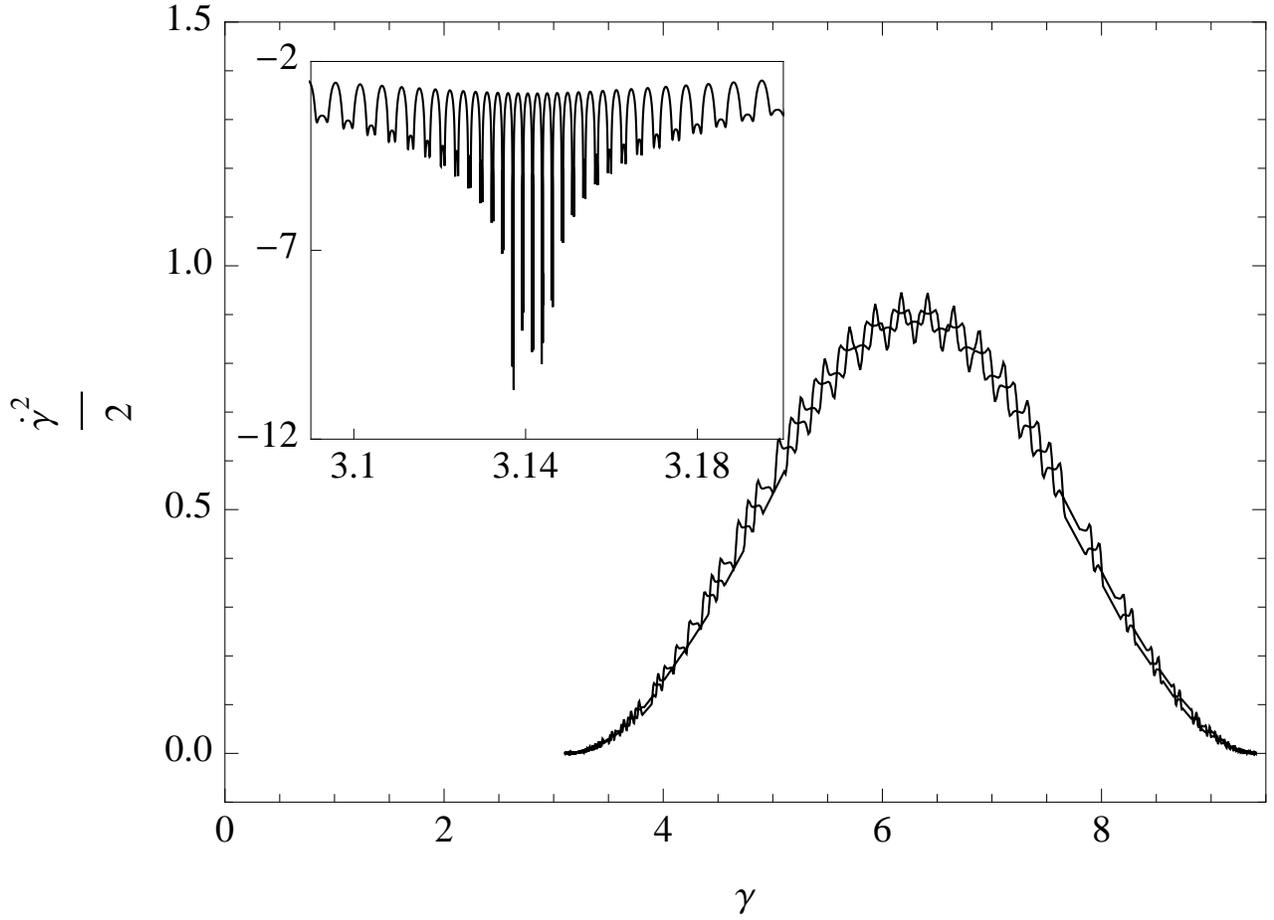}
 \caption{Two librations immediately before and after the point of 2:1 resonance. The resonance is traversed in this case, because the
 descending branch of the second libration never reaches $\,\dot\gamma=0\,$. The inset shows in more detail a fragment of the crucially
 important minimum of the post-resonance libration in logarithmic scale.
 \label{libra.fig}}
 \end{figure}

 Figure \ref{libra.fig} illustrates the role of the oscillatory torques which are ignored in the theoretical calculation of capture
 probability. The graph renders the behaviour of the quantity $\,\dot\gamma/2\,$, which is proportional to the energy of rotation, as a
 function of $\,\gamma\,$ for two full libration cycles, one going forward toward the point of resonance ($\,\dot\gamma>0\,$) and the
 other going back toward $\,\dot\gamma=0\,$ beyond the point of resonance ($\,\dot\gamma<0\,$). Due to the dissipation of energy by the
 tides, the latter curve is systematically lower than the former one, but the difference is so small that it cannot be seen on the graph.
 Besides, the oscillatory terms of the net torque make the curves jittery. Although the amplitude of the jitter diminishes in the
 vicinity of the resonance, it may appear to be significant for the outcome of this process. According to \citet{gol68}, capture occurs
 when the first post-resonance minimum of $\,\dot\gamma^2/2\,$ reaches 0. If the jitter is superimposed with the smooth separatrix
 trajectory, the chances to bump into 0 seem greater than without jitter. The inset in Figure \ref{libra.fig} shows in much greater
 detail this important segment of the post-resonance libration, in the axes $\,\lg(\dot\gamma/2)\,$ versus $\,\gamma\,$. The rotational
 energy comes very close to 0 at the lower extent of oscillations, never quite reaching it. And indeed, in this simulation, the planet
 traversed the 2:1 resonance.

 To get a better understanding of the influence of oscillating terms of tidal torque on the chances of the planet to be captured, we
 repeated the 40 high-accuracy integrations described in the previous paragraph, with the same initial parameters, though this time
 including the entire set of periodic terms. We found that the results for individual simulations often changed, i.e., what resulted in
 capture for a given set of parameters became a passage, and vice versa. Surprisingly, we recorded 14 captures out of 40 trials, yielding
 the same probability of 0.35. From this small-scale experiment, we see that, while the outcome of a particular integration may change
 because of the jitter of the tidal torque, the overall probability of capture is unlikely to depend on it.

 \section{Conclusions}

 The growing number of detected systems of multiple exoplanets and the impressive quality of observational data collected for
 them allow astronomers to perform analysis of probable dynamical states and evolution of these remote worlds at a level of
 detail unthinkable just several years ago. Still, this analysis is riddled with difficulties and uncertainties. The planets
 of GJ 581 present a challenge for both observational practice and interpretational theory. The story of two fictitious planets
 tells a lesson about the hazards of combining, without proper caution, the data from two instruments with their own sets of
 systematic errors. It also calls for a certain unification of the planet detection techniques or, at least, for an easily
 accessible and well-tested standard detection algorithm available as a web application. It is fine to apply a variety of
 different detection methods to the same data, but the standard algorithm should always be checked, and discrepancies, if any,
 should be investigated and reported. $N$-body integration of detected systems should also be a norm reducing the probability
 of error. The system of GJ 581 proves to be remarkably stable, with the four {\it{bona fide}} planets remaining on their
 orbits despite the strong evidence of chaos. The characteristic Lyapunov times are very short compared to the dynamical
 lifetime of the system.

 We have explored the rotation history of the planet GJ 581d assumed to have composition alike to that of the terrestrial
 planets of the Solar system.

 Contrary to the previous publications on the subject, which were based on {\it{ad hoc}}, simplistic tidal theories, we find that this
 planet {\underline{cannot}} be captured into a synchronous or pseudo-synchronous rotation, if it began its evolution from faster,
 prograde initial spin rates. Instead, for a plausible range of parameters, the most likely state of the planet is the 2:1 spin-orbit
 resonance. In this state, the day on GJ 581d should be 67 Earth's days long, which bides for an inhospitable environment, though the
 potential habitability of this planet cannot be ruled out just on climatic considerations. The case of 2:1 spin-orbit resonance was
 considered in the simulations of a hypothetical atmosphere on GJ 581d by \citet{Wordsworth} whose modeling confirmed the possibility
 of liquid water being present on the surface, under some favorite conditions.

 The next likeliest equilibrium states are the 3:2 or 5:2 resonances, depending on the temperature and viscosity of the mantle
 (much less on the planet's triaxiality).

 At the same time, in the event that the initial rotation of the planet was retrograde, the most probable final state is
 synchronous rotation.

 \section*{\large{Acknowledgments}}

 It is our pleasure to thank Jérémy Leconte for a fruitful discussion on the topic of the paper, and for referring us to the work by
 \citet{Wordsworth}.\\

 \section*{\underline{\textbf{\Large{Appendix $\,$A.}}}
 ~~~~~\Large{The tidal torque~~~~~~~~~~~~~~~~}}

 As well known, the tidally generated amendment $\,U\,$ to the potential of the perturbed planet can be presented in the form of a
 Fourier series over the tidal modes
 \ba
 \omega_{lmpq}\;\equiv\;({\it l}-2p)\;\dot{\omega}\,+\,({\it l}-2p+q)\;{\bf{\dot{\cal{M}}}}\,+\,m\;(\dot{\Omega}\,-\,\dot{\theta})
 \,\approx\,(l-2p+q)\,n\,-\,m\,\dot{\theta}\,~,~~~
 \label{L9}
 \ea
 where $\,\dot{\theta}\,$ denotes the rotation rate of the tidally perturbed primary (the planet), while $\,\omega\,$, $\,\Omega\,$,
 $\,n\,$, and $\,{\cal{M}}\,$ are the periapse, the node, the mean motion, and the mean anomaly of the perturbing secondary (the star)
 as seen from the primary. While the tidal modes $\,\omega_{\textstyle{_{lmpq}}}\,$ can be of either sign, the actual physical forcing
 frequencies $\,\chi_{\textstyle{_{lmpq}}}\,$ at which the strain and stress oscillate in the perturbed body are positive-definite:
 \ba
 \chi_{lmpq}\,=~|\,\omega_{lmpq}\,|~\approx~|~(l-2p+q)~n\,-\,m~\dot{\theta}~|\,~.
 \label{fr}
 \ea

 The full series for the tidally generated amendment to the planet's potential was written down by \citet{kaula}, though its partial sum
 was known yet to Sir Charles Darwin (1879). For this reason, we often term this Fourier series as the Darwin-Kaula expansion. We apply
 this term also to the ensuing series for the tidal torque acting on the perturbed planet.

 A detailed derivation of the Fourier expansion of the polar component of the torque can be found in \citet{efrb}. It turns out that the
 torque contains both a secular and a rapidly oscillating part,$\,$\footnote{~The oscillating part averages to
 nil and is not expected to reshuffle much the probabilities of capture (though it can effect the fate of each particular trajectory). It
 can however influence the process of damping of free librations.}$\,$ the secular part being given by
 \ba
 \langle\,{\cal{T}}_z^{\rm{_{\,(TIDE)}}}\rangle\,=\,2\,G\,M_{star}^{{{\,2}}}
 \sum_{{\it{l}}=2}^{\infty}
 \frac{R^{\textstyle{^{2l\,+\,1}}}}{
 a^{\textstyle{^{2l\,+\,2}}}}
 \sum_{m=0}^{l}
 \frac{(l-m)!}{(l+m)!}\;m
 \sum_{p=0}^{l}F^{\,2}_{lmp}(\inc)\sum^{\it \infty}_{q=-\infty}
 G^{\,2}_{lpq}(e)\;k_l(\omega_{lmpq})\;\sin\epsilon_l(\omega_{lmpq})\,~,\quad
 \label{31}
 \ea
 where $\,G\,$ is Newton's gravity constant, $\,a,\,i,\,e\,$ denote the semimajor axis, inclination, and eccentricity, while the
 angular brackets $\,\langle\,.\,.\,.\,\rangle\,$ signify orbital averaging. The standard notations $\,F_{lmp}(\inc)\,$ and $\,G_{lpq}(e)
 \,$ are used for the inclination functions and the eccentricity polynomials. The Love numbers $\,k_{\textstyle{_l}}\,$ and the phase
 lags $\,\epsilon_{\textstyle{_l}}\,$ are functions of the Fourier tidal modes $\,\omega_{\textstyle{_{lmpq}}}\,$ given by (\ref{L9}).

 As explained, e.g., by \citet{e&m}, in the Darwin-Kaula theory of tides the phase lags emerge as the products of the modes
 $\,\omega_{lmpq}\,$ by the appropriate time lags:
 \ba
 \epsilon_l(\omega_{lmpq})\,=\,\omega_{lmpq}~\,\Delta t_l(\omega_{lmpq})\,~,
 \label{lags}
 \ea
 where the lags are written down not as $\,\epsilon_{lmpq}\,$ and $\,\Delta t_{lmpq}\,$ but as $\,\epsilon_l(\omega_{lmpq})\,$ and $\,\Delta
 t_l(\omega_{lmpq})\,$. The same concerns the notation for the Love numbers. This is done in order to emphasise that for a homogeneous
 near-spherical body the functional forms of the lags and Love numbers (as functions of the Fourier mode) are defined solely by the
 integer number $\,l\,$ (called {\it{degree}}).~\footnote{~In his cornerstone work, \citet{kaula} employed the somewhat inconsistent
 notations $\,k_l\,$ and $\,\epsilon_{\textstyle{_{lmpq}}}\,$ which were later borrowed by other authors, e.g., by \citet{efrw} who also
 used a similar notation $\,\Delta t_{lmpq}\,$ for the time lag.
 In our later works \citep{efr1,efrb}, we chose to switch to $\,\epsilon_l\,$ and $\,\Delta t_l\,$, since the forms of the functional
 dependencies of the lags on the modes $\,\omega_{\textstyle{_{lmpq}}}\,$ are defined by the degree $\,l\,$ only. The lags' (and Love
 numbers') dependence on the other three integers, $\,m,\,p,\,q\,$, originates only through the dependence of the argument $\,\omega_{
 \textstyle{_{lmpq}}}\,$ upon these integers. This is why the notations $\,k_{\textstyle{_l}}(\omega_{\textstyle{_{lmpq}}})\,$, $\,
 \epsilon_{\textstyle{_l}}(\omega_{\textstyle{_{lmpq}}})\,$, $\,\Delta t_{\textstyle{_l}}(\omega_{\textstyle{_{lmpq}}})\,$ are more
 adequate than $\,k_{\textstyle{_l}}\,$, $\,\epsilon_{\textstyle{_{lmpq}}}\,$ and $\,\Delta t_{\textstyle{_{lmpq}}}\,$.\\
 $\left.\quad~\right.$ All said applies only to near-spherical homogeneous celestial bodies. For nonspherical bodies, the situation
 becomes more involved, as the functional form of the lags and Love numbers acquires dependence on all the four integers. In this case,
 we should write $\,k_{\textstyle{_{lmpq}}}\,$, $\,\epsilon_{\textstyle{_{lmpq}}}\,$, and $\,\Delta t_{\textstyle{_{lmpq}}}\,$.
 Fortunately, for a slightly non-spherical body, the Love numbers and lags differ from the Love numbers and lags of the spherical
 reference body by terms of the order of the flattening, so a small non-sphericity can be neglected.
 }

 For causality reasons, the time lags $\,\Delta t_{\textstyle{_l}}(\omega_{\textstyle{_{lmpq}}})\,$ are positive-definite. Therefore,
 (\ref{lags}) may be rewritten as
 \ba
 \epsilon_l(\omega_{lmpq})\,=\,\chi_{lmpq}~\,\Delta t_l(\omega_{lmpq})~\,\mbox{Sgn}\,(\,\omega_{lmpq}\,)\,~,
 \label{lags_2}
 \ea
 $\chi_{lmpq}\,$ being the physical forcing frequency (\ref{fr}). As a result of this, the entire expression for the polar component of
 the torque can be written down as
 \ba
 \nonumber
 \langle\,{\cal{T}}_z^{\rm{_{\,(TIDE)}}}\rangle\,=
 ~~~\quad~\quad~\quad~\quad~~\quad~\quad~\quad~\quad~\quad~\quad~\quad~\quad~\quad~\quad~\quad~\quad~\quad~\quad~\quad~\quad~\quad~\quad~\quad~\quad~\quad~\quad~\quad~\quad
 \\
 \nonumber\\
 2\,GM_{star}^{{{\,2}}}
 \sum_{{\it{l}}=2}^{\infty}
 \frac{R^{\textstyle{^{2l\,+\,1}}}}{
 a^{\textstyle{^{2l\,+\,2}}}}
 \sum_{m=0}^{l}
 \frac{(l-m)!}{(l+m)!}\;m
 \sum_{p=0}^{l}F^{\,2}_{lmp}(\inc)\sum^{\it \infty}_{q=-\infty}
 G^{\,2}_{lpq}(e)\;k_l(\omega_{lmpq})~\sin|\,\epsilon_l(\omega_{lmpq})\,|\,~\mbox{Sgn}\,\left(\,\omega_{lmpq}\,\right)
 \,~.\quad
 \label{3111}
 \ea

 When the planet is not too close to the star (${\textstyle R}/{\textstyle a}\ll1$), it is possible to neglect the terms with
 $\,l > 2\,$. For small obliquities ($i\simeq 0$), it is also possible to leave only $\,i-$independent terms (the next-order
 terms being quadratic in $\,i\,$). Finally, for not too large eccentricities ($\,e\ll 1\,$), it is reasonable to take into
 account only the terms up to $\,e^7\,$, inclusive. This would imply
 summation over the values of $\,q\,$ running from $\,-\,7\,$ through $\,7\,$. In reality, however, the term with $\,q\,=\,-\,2\,$ turns
 out to vanish identically, while the terms with $\,q\,=\,-\,7\,$ through $\,q\,=\,-\,3\,$ are accompanied with extremely small numerical
 factors and can thus be dropped. This way, the polar component of the tidal torque gets approximated with
 \ba
 \nonumber
 \langle\,{\cal{T}}_z^{\rm{_{\,(TIDE)}}}\rangle_{\textstyle{_{\textstyle_{\textstyle{_{l=2}}}}}}~=~~\quad~\quad~~\quad~\quad~\quad~\quad
 ~\quad~\quad~\quad~\quad~\quad~\quad~\quad~\quad~\quad~\quad~\quad~\quad~\quad~\quad~\quad~\quad~\quad~\quad~\quad~\quad\\
 \nonumber\\
 \frac{3}{2}~G\,M_{star}^{\,2}\,R^5\,a^{-6}\sum_{q=-1}^{7}\,G^{\,2}_{\textstyle{_{\textstyle{_{20\mbox{\it{q}}}}}}}(e)~k_2(
 \omega_{\textstyle{_{\textstyle{_{220\mbox{\it{q}}}}}}})~\sin|\,\epsilon_2(\omega_{\textstyle{_{\textstyle{_{220\mbox{\it{q}}}}}}})\,|
 \,~\mbox{Sgn}\,\left(\,\omega_{220q}\,\right)+O(e^8\,\epsilon)+O(\inc^2\,\epsilon)~~,~\quad~\quad~
 \label{16b}
 \ea
 Historically, this expression first appeared, without proof, in the paper by \citet{gold} who summed over all integer values of $\,q\,$.
 A schematic proof was later offered by \citet{dobro}.

 The shape
 of the factors $~k_l\,\sin\epsilon_l~$ as functions of the mode $\,\omega_{\textstyle{_{lmpq}}}\,$ is defined by the size and mass of
 the body and by its rheology. A rheological model is a constitutive equation interconnecting the strain and stress. Within linear
 rheologies, such equations can be rewritten in the frequency domain where each harmonic of the strain gets expressed algebraically
 through the appropriate harmonic of the stress. Using the techniques explained in \citet{efr1,efrb}, those algebraic relations can
 be employed to derive the shape of the functions $\,k_l(\omega_{\textstyle{_{lmpq}}})~\sin\epsilon_l(\omega_{\textstyle{_{lmpq}}})\,$
 standing in the terms of the Darwin-Kaula expansion of the tidal torque.

 We rely on the Andrade rheological model which is known to work for Earth's mantle \citep{efr2} and which may therefore be
 applicable to the mantles of other terrestrial planets. The universality of the Andrade model lies also in the fact that it can be
 rewritten in a manner permitting a switch to the Maxwell model at low frequencies \citep{efr1,efrb}. The necessity for
 this switch is dictated by the fact that different physical mechanisms dominate friction over different frequency bands. As demonstrated
 in {\it{Ibid.}}, employment of this combined model (Andrade at higher frequencies, and Maxwell at lower frequencies) renders for
 $\,k_l~\sin\epsilon_l\,$ a kink-shaped dependence upon the Fourier mode -- as the dotted line on Figure \ref{tide.fig}.

 In practical calculations, it is convenient to insert (\ref{L9}) into $~k_l(\omega_{\textstyle{_{lmpq}}})\;\sin\epsilon_l(\omega_{
 \textstyle{_{lmpq}}})~$ and thus to obtain the dependencies of $\,k_l~\sin\epsilon_l\,$ upon the spin rate $\,\dot{\theta}\,$, with
 $\,n\,$ treated as a constant or slow-varying parameter. The dependence of $\,k_l~\sin\epsilon_l\,$ upon $\,\dot{\theta}\,$ will have
 the shape of a kink too. It will be a function varying slowly everywhere except in the vicinity of the spin-orbit resonances $\,\dot{
 \theta} = \frac{\textstyle{l-2p+q}}{\textstyle{m}}\,n~$. The dotted curve in Figure \ref{tide.fig} depicts the
 $\,\dot{\theta}-$dependence of the factor
 \ba
 \nonumber
 k_2(\omega_{\textstyle{_{
 \textstyle{_{2202}}}}})~\sin\epsilon_2(\omega_{\textstyle{_{\textstyle{_{2202}}}}})\,=\,
 k_2(\,4\,n\,-\,2\,\dot{\theta}\,)~\sin\epsilon_2(\,4\,n\,-\,2\,\dot{\theta}\,)\\
 \nonumber\\
 =~k_2(\,4\,n\,-\,2\,\dot{\theta}\,)~~\sin|\,\epsilon_2(\,4\,n\,-\,2\,\dot{\theta}\,)\,|~~\mbox{Sgn}\,(\,4\,n\,-\,2\,\dot{\theta}\,)
 \label{tbt}
 \ea
 in the vicinity of the 2:1 spin-orbit resonance. This factor shows up in the $\,2202\,$ term of the torque. In the 2:1
 spin-orbital resonance, $\,\dot{\theta}\,$ transcends the value of $\,2n\,$, so the tidal mode $\,\omega_{2202} = 2(2n-
 \dot{\theta})\,$ goes through nil and changes its sign. In Figure \ref{tide.fig}, the factor $~k_2(\omega_{\textstyle{_{2202}}})~\sin
 \epsilon_2(\omega_{\textstyle{_{2202}}})~$ does the same: as the tidal mode $\,\omega_{2202}\,$ approaches zero (or, equivalently, as
 $\,\dot{\theta}\,$ goes through $\,2n\,$), the said factor smoothly goes through zero and changes its sign. This makes the considered
 term of the tidal torque change its sign smoothly when the synchronous orbit gets crossed. Note that the rapid but smooth changes of
 tidal torque take place within a very narrow interval of $\,\dot{\theta}\,$. The widely used assumption that the torque component is
 linear in  $\,\dot\theta\,$ (and therefore linear in the tidal mode) is not justified by this model, except in an extremely small range
 of spin rates around the point of resonance.

 Similarly, for any set of integers $\,lmpq\,$, the factor
 \ba
 \nonumber
 k_l(\omega_{\textstyle{_{lmpq}}})\;\sin\epsilon_l(\omega_{\textstyle{_{lmpq}}})~=~k_l(\omega_{\textstyle{_{lmpq}}})\;\sin|\,
 \epsilon_l(\omega_{\textstyle{_{lmpq}}})\,|~\,\mbox{Sgn}\,(\,\omega_{\textstyle{_{lmpq}}}\,)~=~\quad~\quad~\quad~\quad~\quad~\quad~\quad\\
 \nonumber\\
 k_l(\,(l-2p+q)\,n\,-\,m\,\dot{\theta}\,)~~\sin|\,\epsilon_l(\,(l-2p+q)\,n\,-\,m\,\dot{\theta}\,)\,|~~\mbox{Sgn}\,(\,(l-2p+q)\,n\,-\,m
 \,\dot{\theta}\,)\,~,\,\quad
 \label{}
 \ea
 depicted as function of $\,\dot{\theta}\,$ will demonstrate behaviour similar to that of (\ref{tbt}) in Figure \ref{tide.fig}: it will smoothly go
 through nil and will change its sign as the $\,lmpq\,$ commensurability gets transcended, i.e., as the tidal mode
 $\,\omega_{\textstyle{_{lmpq}}}\,=\,(l-2p+q)\,n\,-\,m\,\dot{\theta}\,$ goes through nil.

 As ensues from (\ref{3111}) and (\ref{16b}), the $\,lmpq\,$ term of the torque is decelerating for $\,\omega_{
 \textstyle{_{lmpq}}}<0\,$ and is accelerating for $\,\omega_{\textstyle{_{lmpq}}}>0\,$. This motivates us to use the sign convention as
 in Figure \ref{tide.fig}: the $\,lmpq\,$ term of the torque is agreed to be negative on the right of the $\,lmpq\,$ resonance and positive on its
 left. \footnote{~This sign convention is opposite to the one we used in \citet{efr1,efrb}.}

 The overall tidal torque (\ref{3111}), or its approximation (\ref{16b}), taken as a function of the spin rate $\,\dot{\theta}\,$,
 will look as a superposition of kinks. In other words, if we sum up all the terms in (\ref{3111}) or (\ref{16b}), and depict the sum
 against $\,\dot{\theta}\,$, we shall get an overall curve containing a kink $\,${\it{near}}$\,$ each $\,lmpq\,$ resonance, i.e., $\,${\it{near}}$\,$
 the points $\,\dot{\theta} = \frac{\textstyle{l-2p+q}}{\textstyle{m}}\,n~$. We write $\,${\it{near}}$\,$, because these kinks will not
 go through zero at the said points exactly, but will be slightly displaced. This will happen because the higher-resonance kinks will be
 residing on the slopes of lower-resonance kinks. An example of this is furnished by Figure \ref{tide.fig} where the kink originating
 from the $\,lmpq=2202\,$ term is depicted with the dotted line. The total torque, i.e., the sum of the kink with the bias, is
 given by the solid line. We see that the presence of the bias shifts the kink slightly downward. So the total torque (the solid line)
 crosses the $\,\dot{\theta}/n\,$ axis in a point located slightly to the left of the resonance $\,\dot{\theta}/n = 2\,$.

 \section*{\underline{\textbf{\Large{Appendix $\,$B.}}}
 ~~~~~\Large{How rheology enters the play}}

 A laborious calculation \citep[presented in][]{efr1,efrb} demonstrates that the factors $~k_l(\omega_{\textstyle{_{lmpq}}})~\sin\epsilon_l
 (\omega_{\textstyle{_{lmpq}}})~$ can be expressed via the real and imaginary parts of the complex compliance of the mantle material, and
 the mass and radius of the planet. This way, the mode-dependence of these factors is defined by both the rheology and self-gravitation
 of the planet. The factors come out to be odd functions, which is very natural, since each such factor (and the term of the torque,
 which contains this factor) must change its sign when the appropriate commensurability is transcended.

 Being odd, the factors can then be written down as $~k_l(\omega_{\textstyle{_{lmpq}}})~\sin|\,\epsilon_l(\omega_{\textstyle{_{lmpq}}})\,
 |~\,\mbox{Sgn}\,(\,\omega_{\textstyle{_{lmpq}}}\,)~$, the product $~k_l(\omega_{\textstyle{_{lmpq}}})~\sin|\,\epsilon_l(\omega_{
 \textstyle{_{lmpq}}})\,|~$ being an even function of the tidal mode. In other words, this product may be regarded as a function not of
 the tidal mode $\,\omega_{\textstyle{_{lmpq}}}\,$ but of its absolute value $\,\chi_{\textstyle{_{lmpq}}}\,=\,|\,\omega_{\textstyle{_{
 lmpq}}}\,|\,$, which is the physical forcing frequency of tidal oscillations in the mantle. All in all, we have:
 \ba
 \nonumber
 k_l(\omega_{\textstyle{_{lmpq}}})~\sin\epsilon_l(\omega_{\textstyle{_{lmpq}}})&=&k_l(\omega_{\textstyle{_{lmpq}}})~\sin|\,\epsilon_l
 (\omega_{\textstyle{_{lmpq}}})\,|~\,\mbox{Sgn}\,(\,\omega_{\textstyle{_{lmpq}}}\,)\\
 \nonumber\\
 &=&k_l(\chi_{\textstyle{_{lmpq}}})~\sin|\,\epsilon_l
 (\chi_{\textstyle{_{lmpq}}})\,|~\,\mbox{Sgn}\,(\,\omega_{\textstyle{_{lmpq}}}\,)\,~.
 \label{}
 \ea
  The calculation in {\it{Ibid.}} furnishes the following frequency dependence for these factors:
 \ba
 k_l(\omega_{\textstyle{_{lmpq}}})\;\sin\epsilon_l(\omega_{\textstyle{_{lmpq}}})\,=\;\frac{3}{2\,({\it l}\,-\,1)}\;\,\frac{-\;A_l\;J\;{\cal{I}}{\it{m}}\left[\bar{J}(\chi)\right]}{
 \left(\;{\cal{R}}{\it{e}}\left[\bar{J}(\chi)\right]\;+\;A_l\;J\;\right)^2\;+\;\left(\;{\cal{I}}{\it{m}}\left[\bar{J}(\chi)\right]\;
 \right)^2} ~\,\mbox{Sgn}\,(\,\omega_{\textstyle{_{lmpq}}}\,)~~~,~~~~~
 \label{L39b}
 \ea
 where $\,\chi\,$ is a shortened notation for the frequency $\,\chi_{\textstyle{_{lmpq}}}\,$, while coefficients $\,A_l\,$ are given by
 \ba
 A_{\it l}\,
 \equiv\;\frac{\textstyle{(2\,{\it{l}}^{\,2}\,+\,4\,{\it{l}}\,+\,3)\,{\mu}}}{\textstyle{{\it{l}}\,\mbox{g}\,
 \rho\,R}}\;=\;\frac{\textstyle{3\;(2\,{\it{l}}^{\,2}\,+\,4\,{\it{l}}\,+\,3)\,{\mu}}}{\textstyle{4\;{\it{l}}
 \,\pi\,G\,\rho^2\,R^2}}\;\;\;,~~~~~~~
 \label{}
 \ea
 with $\,R\,$, $\,\rho\,$, $\,\mu\,$, and g  being the radius, mean density, unrelaxed rigidity, and surface gravity of the planet, while
 $\,G\,$ being the Newton gravitational constant.

 The functions $\,{\cal R}{\it e} [\bar{J}(\chi)]\,$ and $\,{\cal I}{\it m} [\bar{J}(\chi)]\,$ are the real and imaginary parts of the
 complex compliance $\,\bar{J}(\chi)\,$ of the mantle. These are rendered by the formulae
 \ba
 {\cal R}{\it e} [ \bar{J}(\chi)]\;=\;J\;+\;J\,(\chi\tau_{_A})^{-\alpha}\;\cos\left(\,\frac{\alpha\,\pi}{2}\,\right)
 \;\Gamma(\alpha\,+\,1)~~~~\quad\quad\quad\quad\quad~\quad\quad\quad\quad\quad\quad\quad
 \label{A4ccc}
 \ea
 and
 \ba
 {\cal I}{\it m} [ \bar{J}(\chi)]\;=\;-\;J~(\chi\tau_{_M})^{-1}\;-\;J\,(\chi\tau_{_A})^{-\alpha}\;\sin\left(
 \,\frac{\alpha\,\pi}{2}\,\right)\;\Gamma(\alpha\,+\,1)~~~,~~~~~~~~~~~~~~\quad\quad\quad
 \label{A3ccc}
 \ea
 $J=1/\mu\,$ being the unrelaxed compliance of the mantle, and $\,\alpha\,$ being a numerical parameter assuming values of about $\,0.3
 \,$ for solid silicates and about $\,0.14 - 0.2\,$ for partial melts. In our computations, we used $\,\alpha=0.2\,$.

 Among the rheological parameters entering (\ref{A4ccc} - \ref{A3ccc}) is the Maxwell time $\,\tau_{_M}\,$, which is the ratio
 of the mantle's viscosity $\,\eta\,$ and rigidity $\,\mu\,$. In the present geological epoch, the Maxwell time of the terrestrial mantle
 is about 500 years. For warmer mantles, it may be much shorter, taken the exponential temperature-dependence of the viscosity.

 Another characteristic time entering the above expressions is the Andrade time $\,\tau_{_A}\,$. Referring the reader to
 \citet{efr1,efrb} for details, we would mention that below some threshold frequency inelastic processes seize to play a major role in the
 internal friction, giving way to viscosity. Thus the mantle's behaviour becomes closer to that of a Maxwell body. Mathematically, this
 means that below the threshold the parameter $\,\tau_{_A}\,$ increases rapidly as the frequency goes down. So only the first term in
 (\ref{A4c}) and the first term in (\ref{A3c}) survive, and we thus are left with the complex compliance of a Maxwell material.

 In our computations, we treated $\,\tau_{_A}\,$ in the same way as in \citet{makti}: we kept $\,\tau_{_A}\,=\,\tau_{_M}\,$ at the
 frequencies above the threshold (which was set to be 1 yr$^{-1}\,$, just like in the solid Earth case). For frequencies lower than that,
 we set $\,\tau_{_A}\,$ to grow exponentially with the decrease of the frequency, so the rheological model approached the Maxwell one in
 the low-frequency limit. Numerical simulation has demonstrated that the resulting capture probabilities are not very sensitive to how
 quickly the switch to the Maxwell model takes place. For more details, see {\it{Ibid}}.

 In a computer code, it is easier to divide both the numerator and denominator of (\ref{L39b}) by $\,J^{\,2}\,$:
 \ba
 k_l(\,\omega_{\textstyle{_{lmpq}}}\,)\;\sin\epsilon_l(\,\omega_{\textstyle{_{lmpq}}}\,)\;=\;\frac{3}{2\,({\it l}\,-\,1)}\;\,\frac{-\;A_l\;{\cal{I}}}{\left(\;{\cal{R}}
 \;+\;A_l\;\right)^2\;+\;{\cal{I}}^{\textstyle{^{\,2}}}}~\,\mbox{Sgn}\,(\,\omega_{\textstyle{_{lmpq}}}\,) ~~~,~~~~~
 \label{L39}
 \ea
 where $\,{\cal R}\,$ and $\,{\cal I}\,$ are the {\it{dimensionless}} real and imaginary parts of the
 complex compliance:
 \ba
 {\cal R}\;=\;1\;+\;(\chi\tau_{_A})^{-\alpha}\;\cos\left(\,\frac{\alpha\,\pi}{2}\,\right)
 \;\Gamma(\alpha\,+\,1)~~~,\quad\quad\quad\quad\quad~\quad\quad\quad\quad\quad\quad\quad
 \label{A4c}
 \ea
 \ba
 {\cal I}\;=\;-\;(\chi\tau_{_M})^{-1}\;-\;(\chi\tau_{_A})^{-\alpha}\;\sin\left(
 \,\frac{\alpha\,\pi}{2}\,\right)\;\Gamma(\alpha\,+\,1)~~~.~~~~~~~~~~~~~~\quad\quad\quad\quad\quad
 \label{A3c}
 \ea
 These dependencies were used also in \citet{makti}, to explore the spin-orbit dynamics of a Mercury-like planet.


 \section*{\underline{\textbf{\Large{Appendix $\,$C.}}}
 ~~~~~\Large{Capture probabilities~~~~~~~~~~}}

 This section offers a squeezed explanation of the capture theory developed by \citet{gold,gol68}.

 As was mentioned in subsection \ref{tt}, and explained at length in Appendix A above, a good approximation for the polar component of
 the tidal torque is
 furnished by expression (\ref{tid.eq}). The tidal-mode-dependent factors entering that expression can be expressed as functions of the
 spin rate $\,\dot{\theta}\,$, see formula (\ref{formula}). Following \citet{gol68}, in the vicinity of each particular resonance $\,q\,
 '\,$, i.e., for $\,\dot{\theta}\,$ being close to $\,n\,\left(1\,+\,{\textstyle q\,'}/{\textstyle 2}\right)\,$, it is instrumental to
 decompose the right-hand side of Equation (\ref{tid.eq}) into two parts, one being the $\,q=q'\,$ kink, another part being the bias. Constituted
 by the inputs from all the $\,q\neq q\,'\,$ modes, the bias is smooth and virtually constant in the vicinity of the $\,q\,'\,$ resonance.

 The $\,q=q'\,$ term is an odd function of $\,\omega_{\textstyle{_{220\mbox{\it{q}}\,'}}}\,$. While \citet{gol68} assumed this term to be
 either a constant multiplied by $\,$Sgn$\,(\omega_{\textstyle{_{220\mbox{\it{q}}\,'}}})\,$ (the constant-torque model) or a constant
 multiplied by the mode $\,\omega_{\textstyle{_{220\mbox{\it{q}}\,'}}}\,$ itself (the linear in frequency model), \citet{makti} endowed
 this term (in fact, all terms) with a realistic mode-dependence originating from the properties of actual rocks.

 It is convenient to introduce an auxiliary variable
 \ba
 \gamma~\equiv~2~\theta~-~(2\,+\,q\,')\,{\cal{M}}
 \label{}
 \ea
 which vanishes when the long axis of the planet points toward the star at the perigee. Then the $\,q=q'\,$ resonance will correspond to
 $\,\dot\gamma=0\,$ because
 \ba
 \dot\gamma\,=\,-~(2\,+\,q\,')\,n\,+\,2\,\dot{\theta}\,\,=\,-\,\omega_{\textstyle{_{220\mbox{\it{q}}\,'}}}\,~.
 \label{}
 \ea
 Be mindful that our $\,\gamma\,$ coincides with that employed by \citet{makti}, and is twice the quantity $\,\gamma\,$ introduced in \citet{gol68}.

 In the absence of tidal friction, the equation of motion near the said resonance looks as
 \ba
 C~\ddot{\gamma} ~+~3~(B\,-\,A)~\frac{M_{star}}{M_{star}\,+\,M_{planet}}~n^2~G_{20\mbox{\small{q}}\,'}(e)~\sin\gamma~=~0\,~.
 \label{*}
 \label{655}
 \ea
 Multiplication thereof by $\,\dot{\gamma}\,$, with subsequent integration over time $\,t\,$, furnishes the first integral of motion,
 \ba
 C~\frac{{\dot{\gamma}}^{\,2}}{2}~-~3~(B\,-\,A)~\frac{M_{star}}{M_{star}\,+\,M_{planet}}~n^2~G_{20\mbox{\small{q}}\,'}(e)~\cos\gamma
 ~=~E\,'~~~,
 \label{**}
 \label{656}
 \ea
 whose value depends on the initial conditions. The latter equation can be rewritten also as
 \ba
 C~\frac{{\dot{\gamma}}^{\,2}}{2}~-~3~(B\,-\,A)~\frac{M_{star}}{M_{star}\,+\,M_{planet}}~n^2~G_{20\mbox{\small{q}}\,'}(e)~
 2~\cos^2\frac{\gamma}{2}~=~E~~~,
 \label{***}
 \label{700}
 \ea
 with $\,E\,$ differing from $\,E'\,$ by a $\,\gamma-$independent constant. As demonstrated in \citet{gol68}, vanishing of the integral
 of motion $\,E\,$ corresponds to the separatrix dividing rotations from librations. So the equation of this separatrix is
 \footnote{~Our Equations (\ref{655}), (\ref{656}), and (\ref{separ.eq}) differ from their counterparts in \citet{gol68}. The difference
 in numerical coefficients originates from our $\,\gamma\,$ being twice that used in {\it{Ibid.}} In our equations, we also keep the mass
 factor omitted in {\it{Ibid}}.}
 \eb
 \dot\gamma~=~2\,n\,\left[\,\frac{3~(B\,-\,A)}{C}~\,\frac{M_{star}}{M_{star}\,+\,M_{planet}}
 ~\,G_{20\mbox{\it{q}\,}'}(e)\,\right]^{{1}/{2}}\cos\frac{\gamma}{2}\,~.
 \label{separ.eq}
 \ee

 In terms of $\,\dot{\gamma}\,$, $\,$the $\,q=q\,'\,$ term of the tidal torque reads as \footnote{~Recall that
 $~k_2(\omega_{\textstyle{_{220\mbox{\it{q}}}}})
 ~\sin|\,\epsilon_2(\omega_{\textstyle{_{220\mbox{\it{q}}}}})\,|~\,\mbox{Sgn}\,(\omega_{\textstyle{_{220\mbox{\it{q}}}}})~$ is an
 odd function, wherefore the switch to the variable $\,\dot{\gamma}\,=\,-\,\omega_{\textstyle{_{220\mbox{\it{q}}\,'}}}\,$ generates
 a ``minus" sign in (\ref{W}).}
 \ba
 W(\dot{\gamma})~=~-~K~G^{\,2}_{\textstyle{_{220\mbox{\it{q}}\,'}}}~k_2(\dot{\gamma})~\sin |\,\epsilon_2(\dot{\gamma})\,|~\,\mbox{Sgn}\,
 (\dot{\gamma})\,~,
 \label{W}
 \ea
 $K\,$ being a positive constant. To write down the bias in the vicinity of this resonance, it is convenient to express an arbitrary
 tidal mode via the resonant one:
 \ba
 \nonumber
 \omega_{\textstyle{_{220\mbox{\it{q}}}}}&=&(2+q)\,n\,-\,2\,\dot{\theta}\,=\,(2\,+\,q\,')\,n\,-\,2\,\dot{\theta}\,+\,(q\,-\,q\,')\,n\\
 \nonumber\\
 &=&-~2\,\dot{\gamma}\,+\,(q\,-\,q\,')\,n\,\approx\,(q\,-\,q\,')\,n\,~,
 \label{13}
 \ea
 where we recalled that $\,\dot{\gamma}\,$ vanishes in the point of resonance. Taking (\ref{13}) into account, we can write the bias as
 \ba
 \nonumber
 V&=&K\,\sum_{q\neq q\,'}G^{\,2}_{220\mbox{\it{q}}}~k_2(\omega_{\textstyle{_{220\mbox{\it{q}}}}})~\sin |\,
 \epsilon_2(\omega_{\textstyle{_{220\mbox{\it{q}}}}})\,|~\,\mbox{Sgn}\,(\omega_{\textstyle{_{220\mbox{\it{q}}}}})\\
 \nonumber\\
 &=&K\,\sum_{q\neq q\,'}G^{\,2}_{220\mbox{\it{q}}}~k_2(\,(q\,-\,q\,')\,n\,)~\sin |\,
 \epsilon_2(\,(q\,-\,q\,')\,n\,)\,|~\,\mbox{Sgn}\,(q\,-\,q\,')\,~,
 \label{V}
 \ea

 For a slowing-down planet, an estimate for the capture probability is derived from the consideration of two librations around the point
 of resonance $\,\dot\gamma\,=\,-\,\omega_{220\mbox{\it{q}\,}'}=0\,$, i.e., of the last libration with a positive $\,\dot\gamma\,$ and the first
 libration with a negative $\,\dot\gamma\,$. \citet{gol68} assumed that the energy offset from zero at the beginning of the last
 libration above the resonance is uniformly distributed between $\,0\,$ and $\,\Delta E=\int\langle T\rangle\dot\gamma dt\,$. This
 assumption rendered them the following estimate for the probability of the capture:
 \eb
 P_{\rm capt}=\frac{\delta E}{\Delta E}\,~,
 \ee
 with $\delta E$ being the total change of the kinetic energy at the end of the libration below the resonance.

 Thus, $\,\langle T\rangle\,\dot\gamma\,$ should be integrated over one cycle of libration, to obtain $\,\Delta E\,$, and should be
 integrated over two librations symmetric around the resonance $\,\omega_{\textstyle{_{220\mbox{\it{q}}\,'}}}=0\,$, to obtain $\,\delta
 E\,$. As a result, the odd part of the tidal torque at $\,q=q\,'\,$ doubles in the integration for $\,\delta E\,$, whereas the bias
 vanishes. Both these components are involved in the computation of $\,\Delta E\,$.

 This makes capture probability look like
 \eb
 P_{\rm capt}\,=\,\frac{2~\int_{-\pi}^{\pi}W(\dot\gamma)\,d\gamma}{\int_{-\pi}^{\pi}\left[\,W(\dot\gamma)\,+\,V\,\right]d\gamma}
 \,\approx\,\frac{2~\int_{-\pi}^{\pi}W(\dot\gamma)\,d\gamma}{\int_{-\pi}^{\pi}W(\dot\gamma)\,d\gamma\,+\,2\,\pi\,V}\,=\,
 \frac{\textstyle 2}{\textstyle{1\,+\,{\textstyle 2~\pi~V}/{\textstyle \int_{-\pi}^{\pi}\,W(\dot\gamma)\,d\gamma}}}\,~.
 \label{prob.eq}
 \ee
 The integral in this equation can be evaluated if we further assume, following \citet{gol68}, that in the vicinity of resonance the
 trajectory follows the singular separatrix solution (\ref{separ.eq}) which corresponds to vanishing of the integral of motion $\,E\,$.

\end{document}